\documentclass[prd,onecolumn,floatfix,nofootinbib，12pt]{revtex4}
\usepackage{amssymb}
\usepackage{amsmath}
\usepackage{graphicx,color,dcolumn,booktabs,bm}
\usepackage{longtable,lscape}
\usepackage{appendix}
\usepackage{txfonts}
\usepackage{overpic}
\usepackage{indentfirst}
\usepackage{cases}
\usepackage{multirow}
\usepackage{epstopdf}
\usepackage{ulem}
\usepackage[colorlinks,
            citecolor=blue,
            anchorcolor=red,
            menucolor=red,
            linkcolor=red,
            filecolor=red,
            runcolor=red,
            urlcolor=blue,
            frenchlinks=false]{hyperref}

\allowdisplaybreaks
\begin{document}

\title{Triply heavy tetraquark states: masses and other properties}
\author{Zhen-Hui Zhu$^{1}$}
\email{ZhuzH88@outlook.com}
\author{Wen-Xuan Zhang$^{1}$}
\email{zhangwx89@outlook.com}
\author{Duojie Jia$^{1,2}$ \thanks{}}
\email{jiadj@nwnu.edu.cn; corresponding author}
\affiliation{$^1$Institute of Theoretical Physics, College of Physics and Electronic
Engineering, Northwest Normal University, Lanzhou 730070, China \\
$^2$Lanzhou Center for Theoretical Physics, Lanzhou University,
Lanzhou,730000,China \\
}

\begin{abstract}
In this work, we study masses and other static properties of triply heavy
tetraquarks in the unified framework of the MIT bag which incorporates
chromomagnetic interactions and enhanced binding energy. The masses,
magnetic moments and charge radii of all strange and nonstrange (ground)
states of triply heavy tetraquarks are computed, suggesting that all of
triply heavy tetraquarks are above the respective two-meson thresholds. We
also estimate relative decay widths of main decay channels of two-heavy
mesons for these tetraquarks. \newline
{\normalsize PACS number(s):12.39Jh, 12.40.Yx, 12.40.Nn}{\normalsize Key
Words: triply heavy tetraquark, masses, magnetic moment and charge radius.}
\end{abstract}

\maketitle
\date{\today }


\section{Introduction}

Study of exotic multi-quark states and hadronic molecules has been an interesting topic and continuous attention since the observation of the
charmoniumlike state $X(3872)$ in 2003 whose quantum number was later shown
to be $J^{P}={1}^{++}$ \cite{PDG:2020}. Since then, a large number of $XYZ$
states were discovered experimentally, such as the charmonium-like states $%
Z_{c}(3885)$ \cite{M. Ablikim et al:2014,LhcXiLf:prl18}, the $Z_{c}(4020)$
\cite{M. Ablikim:2013}, the $Z_{c}(4025)$ \cite{M. Ablikim et al:2014}, the $%
Z_{cs}(3985)$ \cite{M. Ablikim et al:2021}, the $Z_{cs}(4000)$ \cite{R. Aaij
et al:2021}, the $Z_{c}(4200)$ \cite{Belle:2014nuw} and the $Z_{c}(4430)$
\cite{Belle:2013shl}, many of which are possible candidates of multiquark
state or hadronic molecule \cite{Yan:2021tcp,Deng:2021gnb}. Some of the
observed $XYZ$ states, like the charged state $Z_{c}(3900)$ \cite%
{BESIII:2013ris,Belle:2013yex}, are undoubtedly exotic. In 2020, the first
fully charm tetraquark, the $X(6900)$, has been observed by LHCb in the di-$%
J/\Psi $ invariant mass spectrum \cite{LHCb:2020bwg}. In 2021, the first
doubly charmed tetraquark $T_{cc}^{++}$ was discovered in experiment \cite%
{TccPoly:2021}. If these states are confirmed to be exotic, they may be the
cadidates of tetraquarks. Many theoretical discussions exist for the tetraquarks and
multiquarks with different approaches \cite{L.Maiani:2005,Woosung
Park:2014,Muhammad Naeem:2018,Nils A.:1994,Nils A Trnqvist:2004,Eric S.
Swanson:2004,C. Hanhart:2007,F. Aceti:2012,Rui Chen:2016} . For the review
see Refs. (\cite{Shi-Lin Zhu:2005,A. Esposito:2016},\cite%
{Karliner:2017,Eichten:2017,Meng-Lin Du :2013},\cite{Fleck:1989, Ebert:2004,
Roberts:2007, Albertus:2006, Giannuzzi:2009, Bernotas:2008, Liu:2018,
He:2004, KR:2014})

There exist several approaches devoted to triply heavy tetraquark states,
such as AdS/QCD potential\cite{Mutuk:2023yev}, the quark models\cite{Meng:2023jqk,
Liu:2022jdl,Lu:2021kut} and the chromomagnetic interaction model\cite{Guo:2021yws,Cui:2006mp,Xin:2022}. In the discussions, the method of
the MIT bag model is rarely employed to explore triply heavy tetraquarks
though it is successful for light hadrons. Main reason may be that the
conventional MIT bag model fails to apply to heavy hadrons where some extra
binding energy, vanishing between light quarks ($u/d$ and $s$), enters
between heavy quarks($c$ and $b$) and between heavy and strange quarks\cite%
{KR:2014} and are necessary to reconcile light flavor dynamics with heavy
flavor dynamics within an unified bag model\cite{Zhang:2021}. When the chromomagnetic
interaction added which has two different mass scales (one light mass scale
and one heavy mass scale), the strong coupling $\alpha _{s}(R)$ tends to
favor running with the bag radii $R$ in order to correctly describe mass
splittings of hadrons\cite{Zhang:2021}. This makes it involving to explore the heavy
hadrons like heavy tetraquarks in which two heavy quarks(or more) or heavy
and strange quarks are involved using the bag picture like the MIT bag
model.

In this work, we apply the MIT bag model \cite{Zhang:2021,DeGrand:1975}
which incorporate the chromomagnetic interaction and enhanced binding energy
between heavy flavors to calculate the masses, magnetic moments and charge
radii for all ground (strange and nonstrange) states of triply heavy
tetraquarks. The dynamical computation indicates that all of triply heavy
tetraquarks are above the respective two-meson thresholds. By the way, we
estimate the main decay widths of two-heavy mesons for these triply heavy
tetraquarks.

This paper will be organized as follows. In Sect.~\ref{Theory}, we briefly
review the MIT bag model that we use in this work. We present the color-spin
wave functions for tetraquark states in details also in this section. In
Sect.~\ref{tetraquark}, computation is given for the masses and other
properties for the triply heavy tetraquarks. The paper ends with summary in
Sect.~\ref{Discussions}.

\section{The MIT bag model with binding energy}

\label{Theory}

\subsection{Mass formula}

The MIT bag model assumes hadron to be a spherical bag(with the  radius $R$)
with quarks confined in it. The mass formula which, includes the enhanced
binding energy $E_{B}$ and chromomagnetic interactions(CMI) can be written as%
\cite{DeGrand:1975}
\begin{equation}
M\left( R\right) =\sum_{i=1}^{4}\omega _{i}+\frac{4}{3}\pi R^{3}B-\frac{Z_{0}%
}{R}+E_{B}+\langle \Delta H\rangle ,  \label{MBm}
\end{equation}%
\begin{equation}
\omega _{i}=\left( m_{i}^{2}+\frac{x_{i}^{2}}{R^{2}}\right) ^{1/2},
\label{freq}
\end{equation}%
The first three terms contain the kinetic energy of all quarks, the volume
energy with bag constant $B$ and the zero-point energy with coefficient $%
Z_{0}$. The total binding energy $E_{B}=\sum_{I<J}B_{IJ}$ stands for the sum
of the enhanced binding energies $B_{IJ}$ between the $I$ and $%
J$ quarks. The last term $M_{CMI}=\langle \Delta H\rangle $ is the chromomagnetic interaction among quarks within bag, given by
\begin{equation}
M_{CMI}=-\sum_{i<j}\left( \mathbf{\lambda }_{i}\cdot \mathbf{\lambda }%
_{j}\right) \left( \mathbf{\sigma }_{i}\cdot \mathbf{\sigma }_{j}\right)
C_{ij},  \label{CMI}
\end{equation}%
where $\boldsymbol{\lambda }_{i}$ are the Gell-Mann matrices in color space, $%
\boldsymbol{\sigma }_{i}$ are the Pauli matrices in spin space and $C_{ij}$
are the CMI parameters, given by
\begin{equation}
C_{ij}=3\frac{\alpha _{s}\left( R\right) }{R^{3}}\bar{\mu}_{i}\bar{\mu}%
_{j}I_{ij},  \label{Cij}
\end{equation}%
\begin{equation}
\bar{\mu}_{i}=\frac{R}{6}\frac{4\omega _{i}R+2m_{i}R-3}{2\omega _{i}R\left(
\omega _{i}R-1\right) +m_{i}R},  \label{muBari}
\end{equation}%
\begin{equation}
I_{ij}=1+2\int_{0}^{R}\frac{dr}{r^{4}}\bar{\mu}_{i}\bar{\mu}%
_{j}=1+F(x_{i},x_{j}).  \label{Iij}
\end{equation}%
Here, $\bar{\mu}_{i}$ is reduced magnetic moment, $F(x_{i},x_{j})$ a
rational function and $\alpha (R)$ the strong coupling, all of which are
detailed in Appendix A, where two relations  relations for $F(x_{i},x_{j})$ and $\alpha (R)$ of computing the CMI matrix
elements is also given.

The binding energy $B_{IJ}$ stems mainly from the short-range chromoelectric
interaction between heavy quark $I$ and/or antiquark $J$, which is greatly
enhanced compared to that between light quarks. Since the latter is much
smaller, it can be ignored. In the case that the quark ($I$) is heavy and the other quark ($J$) is strange, we assume $B_{IJ}$ to be
nonvanishing, as assumed in Ref. \cite{KR:2014,Zhang:2021}. As such, 
 for the color configuration $\boldsymbol{\bar{3}}_{c}$ of the quark pair, there exist five binding energies :
$B_{cs}$, $B_{cc}$, $B_{bs}$, $B_{bb}$ and $B_{bc}$.

For the other parameters like quark masses, zero-point energy $Z_{0}$ in Eqs. (\ref{MBm}) and (\ref{freq}) except for $B_{IJ}$ , we apply the following
input values
\cite{DeGrand:1975,Zhang:2021}. For the binding energy between quark pairs $ij$ in the rep.$\boldsymbol{\bar{3}}_{c}$ , we employ the values from Ref. \cite{Zhang:2021}:
\begin{equation}
\begin{Bmatrix}
m_{n}=0, & m_{s}=0.279\,\text{GeV,} & m_{c}=1.641\text{GeV,} & m_{b}=5.093%
\text{GeV,} \\
Z_{0}=1.83, & B^{1/4}=0.145\,\text{GeV,} & \alpha _{s}=0.55. &
\end{Bmatrix}
\label{oripar}
\end{equation}

\begin{equation}
\begin{Bmatrix}
B_{cs}=-0.025\,\text{GeV,} & B_{cc}=-0.077\,\text{GeV,} & m_{bs}=-0.032\,%
\text{GeV,} \\
B_{bb}=-0.128\,\text{GeV,} & B_{bc}=-0.101\,\text{GeV.} &
\end{Bmatrix}%
\end{equation}

In computation, we solve the transcendental equation(\ref{transc})
interactively to find $x_{i}$ depending on $R$. Here, the bag radii $R$
varies with hadrons and can be solved numerically by Eq. (\ref{MBm}) using
variational method provided that the CMI matrix is evaluated for a given
configuration of tetraquark wavefunction, which we address in the following.

\subsection{The color-spin wavefunction.}

\label{wave function} In this subsection we present the color-spin
wavefunctions of all triply heavy tetraquark states which are required to
evaluate the CMI mass splitting due to the chromommagnetic interaction in
Eq. (\ref{CMI}). For simplicity, we express the flavor parts of the
tetraquarks explicitly in terms of the quark flavors($n=u/d$, $s$, $c$ and $%
b $). For the colorless tetraquark $T$, its color wavefunction has two color
structures $\boldsymbol{6}_{c}\otimes \boldsymbol{\bar{6}}_{c}$ or $%
\boldsymbol{\bar{3}}_{c}\otimes \boldsymbol{3}_{c}$ in the notation of the
color representations $3_{c}(\bar{3}_{c})$ and $6_{c}(\bar{6}_{c})$,
corresponding to the following color configurations
\begin{equation}
\phi _{1}^{T}=\left\vert {\left( q_{1}q_{2}\right) }^{6}{\left( \bar{q}_{3}%
\bar{q}_{4}\right) }^{\bar{6}}\right\rangle ,\quad \phi _{2}^{T}=\left\vert {%
\left( q_{1}q_{2}\right) }^{\bar{3}}{\left( \bar{q}_{3}\bar{q}_{4}\right) }%
^{3}\right\rangle ,  \label{colorT}
\end{equation}%
respectively. For the spin part of tetraquark wavefunction, there are six
configurations (spin bases in spin space)
\begin{equation}
\begin{aligned} \chi_{1}^{T}={\left| {\left(q_{1}q_{2}\right)}_{1}
{\left(\bar{q}_{3}\bar{q}_{4}\right)}_{1} \right\rangle}_{2}, \quad
\chi_{2}^{T}={\left| {\left(q_{1}q_{2}\right)}_{1}
{\left(\bar{q}_{3}\bar{q}_{4}\right)}_{1} \right\rangle}_{1}, \\
\chi_{3}^{T}={\left| {\left(q_{1}q_{2}\right)}_{1}
{\left(\bar{q}_{3}\bar{q}_{4}\right)}_{1} \right\rangle}_{0}, \quad
\chi_{4}^{T}={\left| {\left(q_{1}q_{2}\right)}_{1}
{\left(\bar{q}_{3}\bar{q}_{4}\right)}_{0} \right\rangle}_{1}, \\
\chi_{5}^{T}={\left| {\left(q_{1}q_{2}\right)}_{0}
{\left(\bar{q}_{3}\bar{q}_{4}\right)}_{1} \right\rangle}_{1}, \quad
\chi_{6}^{T}={\left| {\left(q_{1}q_{2}\right)}_{0}
{\left(\bar{q}_{3}\bar{q}_{4}\right)}_{0} \right\rangle}_{0}, \end{aligned}
\label{spinT}
\end{equation}

Combining the color and spin parts of wavefunctions, tetraquark has twelve
wavefunctions in their ground($1S$) state,
\begin{equation}
\begin{aligned} \phi_{1}^{T}\chi_{1}^{T}={\left|
{\left(q_{1}q_{2}\right)}_{1}^{6}
{\left(\bar{q}_{3}\bar{q}_{4}\right)}_{1}^{\bar{6}} \right\rangle}_{2}
\delta_{12}^{A}\delta_{34}^{A}, \\ \phi_{2}^{T}\chi_{1}^{T}={\left|
{\left(q_{1}q_{2}\right)}_{1}^{\bar{3}}
{\left(\bar{q}_{3}\bar{q}_{4}\right)}_{1}^{3} \right\rangle}_{2}
\delta_{12}^{S}\delta_{34}^{S}, \\ \phi_{1}^{T}\chi_{2}^{T}={\left|
{\left(q_{1}q_{2}\right)}_{1}^{6}
{\left(\bar{q}_{3}\bar{q}_{4}\right)}_{1}^{\bar{6}} \right\rangle}_{1}
\delta_{12}^{A}\delta_{34}^{A}, \\ \phi_{2}^{T}\chi_{2}^{T}={\left|
{\left(q_{1}q_{2}\right)}_{1}^{\bar{3}}
{\left(\bar{q}_{3}\bar{q}_{4}\right)}_{1}^{3} \right\rangle}_{1}
\delta_{12}^{S}\delta_{34}^{S}, \\ \phi_{1}^{T}\chi_{3}^{T}={\left|
{\left(q_{1}q_{2}\right)}_{1}^{6}
{\left(\bar{q}_{3}\bar{q}_{4}\right)}_{1}^{\bar{6}} \right\rangle}_{0}
\delta_{12}^{A}\delta_{34}^{A}, \\ \phi_{2}^{T}\chi_{3}^{T}={\left|
{\left(q_{1}q_{2}\right)}_{1}^{\bar{3}}
{\left(\bar{q}_{3}\bar{q}_{4}\right)}_{1}^{3} \right\rangle}_{0}
\delta_{12}^{S}\delta_{34}^{S}, \\ \phi_{1}^{T}\chi_{4}^{T}={\left|
{\left(q_{1}q_{2}\right)}_{1}^{6}
{\left(\bar{q}_{3}\bar{q}_{4}\right)}_{0}^{\bar{6}} \right\rangle}_{1}
\delta_{12}^{A}\delta_{34}^{S}, \\ \phi_{2}^{T}\chi_{4}^{T}={\left|
{\left(q_{1}q_{2}\right)}_{1}^{\bar{3}}
{\left(\bar{q}_{3}\bar{q}_{4}\right)}_{0}^{3} \right\rangle}_{1}
\delta_{12}^{S}\delta_{34}^{A}, \\ \phi_{1}^{T}\chi_{5}^{T}={\left|
{\left(q_{1}q_{2}\right)}_{0}^{6}
{\left(\bar{q}_{3}\bar{q}_{4}\right)}_{1}^{\bar{6}} \right\rangle}_{1}
\delta_{12}^{S}\delta_{34}^{A}, \\ \phi_{2}^{T}\chi_{5}^{T}={\left|
{\left(q_{1}q_{2}\right)}_{0}^{\bar{3}}
{\left(\bar{q}_{3}\bar{q}_{4}\right)}_{1}^{3} \right\rangle}_{1}
\delta_{12}^{A}\delta_{34}^{S}, \\ \phi_{1}^{T}\chi_{6}^{T}={\left|
{\left(q_{1}q_{2}\right)}_{0}^{6}
{\left(\bar{q}_{3}\bar{q}_{4}\right)}_{0}^{\bar{6}} \right\rangle}_{0}
\delta_{12}^{S}\delta_{34}^{S}, \\ \phi_{2}^{T}\chi_{6}^{T}={\left|
{\left(q_{1}q_{2}\right)}_{0}^{\bar{3}}
{\left(\bar{q}_{3}\bar{q}_{4}\right)}_{0}^{3} \right\rangle}_{0}
\delta_{12}^{A}\delta_{34}^{A}. \end{aligned}  \label{colorspinT}
\end{equation}%
where the $\delta $ is defined to take flavor symmetry into account when
writting tetraquark wavefunction. Firstly, $\delta _{12}^{S}=0$($\delta
_{12}^{A}=0$) if two quarks $q_{1}$ and $q_{2}$ are antisymmetric(symmetric)
in flavor space; Secondly, $\delta _{34}^{A}=0$($\delta _{34}^{S}=0$) if $%
q_{3}$ and $q_{4}$ are symmetric(antisymmetric) in flavor space; Thirdly, $%
\delta _{12}^{S}=\delta _{12}^{A}=1$ and $\delta _{34}^{S}=\delta
_{34}^{A}=1 $ in all other cases.

For given color configurations one can write binding energy explicitly $%
E_{B}=\sum_{I<J}B_{IJ}=\sum_{I<J}g_{IJ}B_{IJ}(\bar{3}_{c})$ with the help of
the ratios $g_{IJ}$ of the color factor $\langle \mathbf{\lambda }_{I}\cdot
\mathbf{\lambda }_{J}\rangle $ for given quark pair $q_{I}q_{J}$ relative to
the color factor $\langle \mathbf{\lambda }_{I}\cdot \mathbf{\lambda }%
_{J}\rangle _{\bar{3}_{c}}$ for the quark pair $q_{I}q_{J}$ in the color
antitriplet($\bar{3}_{c}$), provided that the binding energy linearly scales
with this factor. In the case of quark pairs in the color antitriplet $\bar{3%
}_{c}$ and singlet $1_{c}$, this is known as the $1/2$-rule the short-range
interaction: $B([qq^{\prime }]_{3_{c}})=(1/2)B([qq^{\prime }]_{1_{c}})$. For
the tetraquark with color configuration $\phi _{1}^{T}=\left\vert {\left(
q_{1}q_{2}\right) }^{6}{\left( \bar{q}_{3}\bar{q}_{4}\right) }^{\bar{6}%
}\right\rangle $, one can compute scaling ratios $g_{IJ}$\cite{Zhang:2021}
for each pair $q_{I}q_{J}$ in tetraquark ${\left( q_{1}q_{2}\right) \left(
\bar{q}_{3}\bar{q}_{4}\right) }$ and find the scaled binding energy $%
g_{IJ}B_{IJ}(\bar{3}_{c})$, giving rised to total binding energy,
\begin{equation}
E_{B}(\phi _{1}^{T})=-\frac{1}{2}B_{12}+\frac{5}{4}B_{13}+\frac{5}{4}B_{14}+%
\frac{5}{4}B_{23}+\frac{5}{4}B_{24}-\frac{1}{2}B_{34}.  \label{ph1binding}
\end{equation}%
Similarly, for the tetraquark with color configuration $\phi
_{2}^{T}=\left\vert {\left( q_{1}q_{2}\right) }^{\bar{3}}{\left( \bar{q}_{3}%
\bar{q}_{4}\right) }^{3}\right\rangle $, the total binding energy is found
to be\cite{Zhang:2021}
\begin{equation}
E_{B}(\phi _{2}^{T})=B_{12}+\frac{1}{2}B_{13}+\frac{1}{2}B_{14}+\frac{1}{2}%
B_{23}+\frac{1}{2}B_{24}+B_{34},  \label{ph2binding}
\end{equation}%
where $B_{ij}=B_{ij}(\bar{3}_{c})$ stands for the enhanced binding energy
between quark $i$ and $j$ in $\bar{3}_{c}$, which are nonvanishing only
between heavy quarks or between heavy quarks and strange quarks. Here, the
numeric prefactors in front of $B_{ij}$ are the color factor ratios $g_{ij}$
of the quark pair ($i,j$) in given configurations $\phi _{1,2}^{T}$. In the
case of the quark pair $12$ in $6_{c}$ in the state $\phi _{1}^{T}$, for
instance, the color factor ratio is $-1:2$ relative to that in $\bar{3}_{c}$%
, and that for the pair $13$ the ratio is $5:4$ relative to that in $\bar{3}%
_{c}$. For the details of the color factors and their ratios \cite%
{Zhang:2021}, see Appendix B.

Notice that in the color representation $\bar{3}_{c}$, the binding energy
for the given flavor of the quark pairs $ij$ can be obtained from Ref. \cite%
{Zhang:2021}.

For the color or spin matrices between quarks in given tetraquark with
wavefunctions $\phi _{1,2}^{T}$, one can employ the respective matrix
formula (\ref{colorfc}) and (\ref{spinfc}) in Appendix B. For tetraquarks
with wavefunction bases $(\phi _{1}^{T},\phi _{2}^{T})$, the color factor
matrices are given explicitly in Appendix B.

Using the CG coefficients for the given tetraquark configurations $\phi
_{1,2}^{T}$, one can write the six spin wavefunctions explicitly. They are
\begin{equation}
\begin{aligned} \chi_{1}^{T}&=\uparrow\uparrow\uparrow\uparrow, \\
\chi_{2}^{T}&=\frac{1}{2}
\left(\uparrow\uparrow\uparrow\downarrow+\uparrow\uparrow\downarrow\uparrow-%
\uparrow\downarrow\uparrow\uparrow-\downarrow\uparrow\uparrow\uparrow%
\right), \\ \chi_{3}^{T}&=\frac{1}{\sqrt{3}}
\left(\uparrow\uparrow\downarrow\downarrow+\downarrow\downarrow\uparrow%
\uparrow\right), \\ &-\frac{1}{2\sqrt{3}}
\left(\uparrow\downarrow\uparrow\downarrow+\uparrow\downarrow\downarrow%
\uparrow+\downarrow\uparrow\uparrow\downarrow+\downarrow\uparrow\downarrow%
\uparrow\right) \\ \chi_{4}^{T}&=\frac{1}{\sqrt{2}}
\left(\uparrow\uparrow\uparrow\downarrow-\uparrow\uparrow\downarrow\uparrow%
\right), \\ \chi_{5}^{T}&=\frac{1}{\sqrt{2}}
\left(\uparrow\downarrow\uparrow\uparrow-\downarrow\uparrow\uparrow\uparrow%
\right), \\ \chi_{6}^{T}&=\frac{1}{2}
\left(\uparrow\downarrow\uparrow\downarrow-\uparrow\downarrow\downarrow%
\uparrow-\downarrow\uparrow\uparrow\downarrow+\downarrow\uparrow\downarrow%
\uparrow\right), \end{aligned}  \label{pspinT}
\end{equation}%
where the notation uparrow stands for the state of spin upward, and the
downarrow for the state of spin downwards. Given the spin wavefunctions (\ref%
{pspinT}), one can then compute the spin matrices $\langle \mathbf{\sigma }%
_{i}\cdot \mathbf{\sigma }_{j}\rangle $ of tetraquark states via the
formulas (\ref{spinfc}) in Appendix B. There are six spin matrices for the
tetraquark in the subspace of $\{\chi _{1-6}^{T}\}$, which are
\begin{align}
\left\langle \boldsymbol{\sigma _{1}}\cdot \boldsymbol{\sigma _{2}}%
\right\rangle & =%
\begin{bmatrix}
1 & 0 & 0 & 0 & 0 & 0 \\
0 & 1 & 0 & 0 & 0 & 0 \\
0 & 0 & 1 & 0 & 0 & 0 \\
0 & 0 & 0 & 1 & 0 & 0 \\
0 & 0 & 0 & 0 & -3 & 0 \\
0 & 0 & 0 & 0 & 0 & -3%
\end{bmatrix}%
, \\
\left\langle \boldsymbol{\sigma _{1}}\cdot \boldsymbol{\sigma _{3}}%
\right\rangle & =%
\begin{bmatrix}
1 & 0 & 0 & 0 & 0 & 0 \\
0 & -1 & 0 & \sqrt{2} & -\sqrt{2} & 0 \\
0 & 0 & -2 & 0 & 0 & -\sqrt{3} \\
0 & \sqrt{2} & 0 & 0 & 1 & 0 \\
0 & -\sqrt{2} & 0 & 1 & 0 & 0 \\
0 & 0 & -\sqrt{3} & 0 & 0 & 0%
\end{bmatrix}%
, \\
\left\langle \boldsymbol{\sigma _{1}}\cdot \boldsymbol{\sigma _{4}}%
\right\rangle & =%
\begin{bmatrix}
1 & 0 & 0 & 0 & 0 & 0 \\
0 & -1 & 0 & -\sqrt{2} & -\sqrt{2} & 0 \\
0 & 0 & -2 & 0 & 0 & \sqrt{3} \\
0 & -\sqrt{2} & 0 & 0 & -1 & 0 \\
0 & -\sqrt{2} & 0 & -1 & 0 & 0 \\
0 & 0 & \sqrt{3} & 0 & 0 & 0%
\end{bmatrix}%
, \\
\left\langle \boldsymbol{\sigma _{2}}\cdot \boldsymbol{\sigma _{3}}%
\right\rangle & =%
\begin{bmatrix}
1 & 0 & 0 & 0 & 0 & 0 \\
0 & -1 & 0 & \sqrt{2} & \sqrt{2} & 0 \\
0 & 0 & -2 & 0 & 0 & \sqrt{3} \\
0 & \sqrt{2} & 0 & 0 & -1 & 0 \\
0 & \sqrt{2} & 0 & -1 & 0 & 0 \\
0 & 0 & \sqrt{3} & 0 & 0 & 0%
\end{bmatrix}%
, \\
\left\langle \boldsymbol{\sigma _{2}}\cdot \boldsymbol{\sigma _{4}}%
\right\rangle & =%
\begin{bmatrix}
1 & 0 & 0 & 0 & 0 & 0 \\
0 & -1 & 0 & -\sqrt{2} & \sqrt{2} & 0 \\
0 & 0 & -2 & 0 & 0 & -\sqrt{3} \\
0 & -\sqrt{2} & 0 & 0 & 1 & 0 \\
0 & \sqrt{2} & 0 & 1 & 0 & 0 \\
0 & 0 & -\sqrt{3} & 0 & 0 & 0%
\end{bmatrix}%
, \\
\left\langle \boldsymbol{\sigma _{3}}\cdot \boldsymbol{\sigma _{4}}%
\right\rangle & =%
\begin{bmatrix}
1 & 0 & 0 & 0 & 0 & 0 \\
0 & 1 & 0 & 0 & 0 & 0 \\
0 & 0 & 1 & 0 & 0 & 0 \\
0 & 0 & 0 & -3 & 0 & 0 \\
0 & 0 & 0 & 0 & 1 & 0 \\
0 & 0 & 0 & 0 & 0 & -3%
\end{bmatrix}
\label{sfcT}
\end{align}

\subsection{Properties of triply heavy tetraquark states}

In the framework of MIT bag model, the masses, magnetic moments and charge
radii can all be calculated for the ground state hadrons, as shown in sect.
2.A. Given the model parameters and formulas in section 2, one can then
compute the masses of the triply heavy tetraquark states. We review in this
subsection the related relations for other properties (magnetic moments and
charge radii) of triply heavy tetraquark states.

Following the method in bag model\cite{Chodos:1974}, the contribution of a
quark $i$ or an antiquark $i$ with electric charge $Q_{i}$ to the charge
radii of hadron is
\begin{equation}
\begin{aligned} {\left\langle r_{E}^{2} \right\rangle}_{i}&= Q_{i}R^{2}
\frac{\alpha_{i}
\left[2x_{i}^{2}\left(\alpha_{i}-1\right)+4\alpha_{i}+2\lambda_{i}-3\right]}
{3x_{i}^{2} \left[2\alpha_{i}\left(\alpha_{i}-1\right)+\lambda_{i}\right]}
\\ &-Q_{i}R^{2} \frac{\lambda_{i}
\left[4\alpha_{i}+2\lambda_{i}-2x_{i}^{2}-3\right]} {2x_{i}^{2}
\left[2\alpha_{i}\left(\alpha_{i}-1\right)+\lambda_{i}\right]}. \end{aligned}
\label{rEi}
\end{equation}%
Summing all of these contributions in Eq. (\ref{rEi}) gives rise to the
charge radii of hadron\cite{Chodos:1974},
\begin{equation}
r_{E}={\left\vert \sum\nolimits_{i}{\left\langle r_{E}^{2}\right\rangle }%
_{i}\right\vert }^{1/2}.  \label{rEsum}
\end{equation}%
Note that Eq. (\ref{rEsum}) also holds true when the system has the
identical quark constituents so that chromomagnetic interaction may yield
spin-color state mixing.

For the magnetic moment, the related formula is\cite{Chodos:1974,Zhang:2021}%
.
\begin{equation}
\mu _{i}=Q_{i}\bar{\mu}_{i}=Q_{i}\frac{R}{6}\frac{4\omega _{i}R+2m_{i}R-3}{%
2\omega _{i}R\left( \omega _{i}R-1\right) +m_{i}R},  \label{mui}
\end{equation}%
\begin{equation}
\mu =\left\langle \psi _{spin}\left\vert \sum\nolimits_{i}g_{i}\mu
_{i}S_{iz}\right\vert \psi _{spin}\right\rangle ,  \label{musum}
\end{equation}%
where $g_{i}=2$ and $S_{iz}$ is the third component of quark $i$ and
antiquark $i$.

\renewcommand{\tabcolsep}{2.5cm} \renewcommand{\arraystretch}{1.5}
\begin{table}[tbh]
\caption{Sum rule for magnetic moments of tetraquarks $\left(
q_{1}q_{2}\right) \left( {\bar{q}}_{3}{\bar{q}}_{4}\right) $ and their
spin-mixed systems.}
\label{tab:musum}%
\begin{tabular}{cc}
\hline\hline
$\psi_{spin}$ & $\mu$ \\ \hline
$\chi_{1}^{T}$ & $\mu_{1}+\mu_{2}+\mu_{3}+\mu_{4}$ \\
$\chi_{2}^{T}$ & $\frac{1}{2}\left(\mu_{1}+\mu_{2}+\mu_{3}+\mu_{4}\right)$
\\
$\chi_{3}^{T}$ & 0 \\
$\chi_{4}^{T}$ & $\mu_{1}+\mu_{2}$ \\
$\chi_{5}^{T}$ & $\mu_{3}+\mu_{4}$ \\
$\chi_{6}^{T}$ & 0 \\
$C_{1}\chi_{3}^{T}+C_{2}\chi_{6}^{T}$ & 0 \\ \hline\hline
\end{tabular}%
\end{table}
We list in the Table \ref{tab:musum} all magnetic moment formulas as a sum
rule for the triply heavy tetraquarks. Given spin wavefunction $\chi
_{1-6}^{T}$, the corresponding magnetic moment of triply heavy tetraquark
states can be calculated via Table I. In the case of spin mixing, the
computation goes in the following way: The spin wavefunction $C_{1}\chi
_{2}^{T}+C_{2}\chi _{4}^{T}+C_{3}\chi _{5}^{T}$ contributes to magnetic
moment by $C_{1}^{2}\mu \left( \chi _{2}^{T}\right) +C_{2}^{2}\mu \left(
\chi _{4}^{T}\right) +C_{3}^{2}\mu \left( \chi _{5}^{T}\right) +\sqrt{2}%
C_{1}C_{2}\left( \mu _{3}-\mu _{4}\right) +\sqrt{2}C_{1}C_{3}\left( \mu
_{2}-\mu _{1}\right) $, the spin wave function $C_{1}\chi _{2}^{T}+C_{2}\chi
_{5}^{T}+C_{3}\chi _{4}^{T}$ contributes $C_{1}^{2}\mu \left( \chi
_{2}^{T}\right) +C_{2}^{2}\mu \left( \chi _{5}^{T}\right) +C_{3}^{2}\mu
\left( \chi _{4}^{T}\right) +\sqrt{2}C_{1}C_{2}\left( \mu _{2}-\mu
_{1}\right) +\sqrt{2}C_{1}C_{3}\left( \mu _{3}-\mu _{4}\right) $.

In addition, we present the way to compute the decay width $\Gamma _{I}$ of
the tetraquark state\cite{Xin:2022}, which is given by
\begin{equation}
\Gamma _{I}=\gamma _{I}\alpha {\frac{p^{2L+1}}{m^{2L}}\cdot {\left\vert {%
c_{I}}\right\vert }^{2}},  \label{decy}
\end{equation}%
with $\gamma _{I}$ the prefactor involving deacy dynamics, $\alpha $ a
coupling constant at decay vertex, $p$ is the momentum of final hadrons(two
mesons $M_{1}$ and $M_{2}$), $m$ the mass of inital tetraquark state and $%
c_{I}$ eigenvector of the $I$-th decay channel. The factor $\gamma _{I}$
varies depending upon the initial tetraquark and the final decay state and
it differs for each decay channel. Notice that the following prefactors are
nearly same
\begin{equation}
\gamma _{M_{1}M_{2}}=\gamma _{{M_{1}}^{\ast }M_{2}}=\gamma _{M_{1}{M_{2}}%
^{\ast }}=\gamma _{{M_{1}}^{\ast }{M_{2}}^{\ast }}  \label{M1}
\end{equation}%
for the final channel $M_{1}M_{2}$, ${M_{1}}^{\ast }M_{2}$, $M_{1}{M_{2}}%
^{\ast }$ and ${M_{1}}^{\ast }{M_{2}}^{\ast }$ where $M_{1,2}$ and ${M_{1,2}}%
^{\ast }$ stand for the mesons or excited mesons in the final decay states,
respectively.

\section{NUMERICAL RESULTS}

\label{tetraquark}

Given the model parameters and formulas detailed in section 2, one can
compute the masses, magnetic moments and charge radii of all ground
(strange and nonstrange) states of triply heavy tetraquarks. For this
purpose, we proceed this computation for the tetraquarks with three heavy
quarks(antiquarks) by grouping them into four classes, namely, the triply
heavy tetraquarks $nb\bar{Q}\bar{Q}$ and $nc\bar{Q}\bar{Q}$, the triply
heavy tetraquarks $sc\bar{Q}\bar{Q}$, $sb\bar{Q}\bar{Q}$, the triply
heavy tetraquarks $nQ\bar{Q}\bar{Q\prime }$ and the triply
heavy tetraquarks $sQ\bar{Q}\bar{Q\prime }$, as to be addressed in the
following subsections. We use the notation $T(qQ^{\prime \prime }\bar{Q}\bar{%
Q}^{\prime },M,J^{P})$ denote the triply heavy tetraquark with flavor
content $qQ^{\prime \prime }\bar{Q}\bar{Q}^{\prime }$, mass $M$ in GeV and
the spin-parity $J^{P}$, for simplicity.

\subsection{Triply heavy tetraquarks $nb\bar{Q}\bar{Q}$ and $nc\bar{Q}\bar{Q}
$.}

We consider first the triply heavy tetraquarks $nb\bar{Q}\bar{Q}$ and $nc%
\bar{Q}\bar{Q}$($Q=c,b$) with identical flavor of two heavy antiquark. Owing
to the chromomagnetic interaction (\ref{CMI}), the color-spin states ($\phi
\chi $) of tetraquark $nb\bar{Q}\bar{Q}$ or $nc\bar{Q}\bar{Q}$ can mix among
them to form a mixed state, a linear superposition of spin-color bases. For
the heavy tetraquark, the $J^{P}=2^{+}$ state consists of one state ($\phi
_{2}\chi _{1}$), with no mixing. The $J^{P}=1^{+}$ state consists of three
basis states ($\phi _{2}\chi _{2},\phi _{2}\chi _{5},\phi _{1}\chi _{4}$)
and can mix into a mixed state having the form of $c_{1}\phi _{2}\chi
_{2}+c_{2}\phi _{2}\chi _{5}+c_{3}\phi _{1}\chi _{4}$. Meanwhile, the $%
J^{P}=0^{+}$ state, consisting of two basis states ($\phi _{2}\chi _{3},\phi
_{1}\chi _{6}$), can mix into the mixed states with the form $c_{1}\phi
_{2}\chi _{3}+c_{2}\phi _{1}\chi _{6}$.

\begin{figure}[tph]
\setlength{\abovecaptionskip}{0.1cm}
\par
\begin{center}
\includegraphics[width=0.5\textwidth]{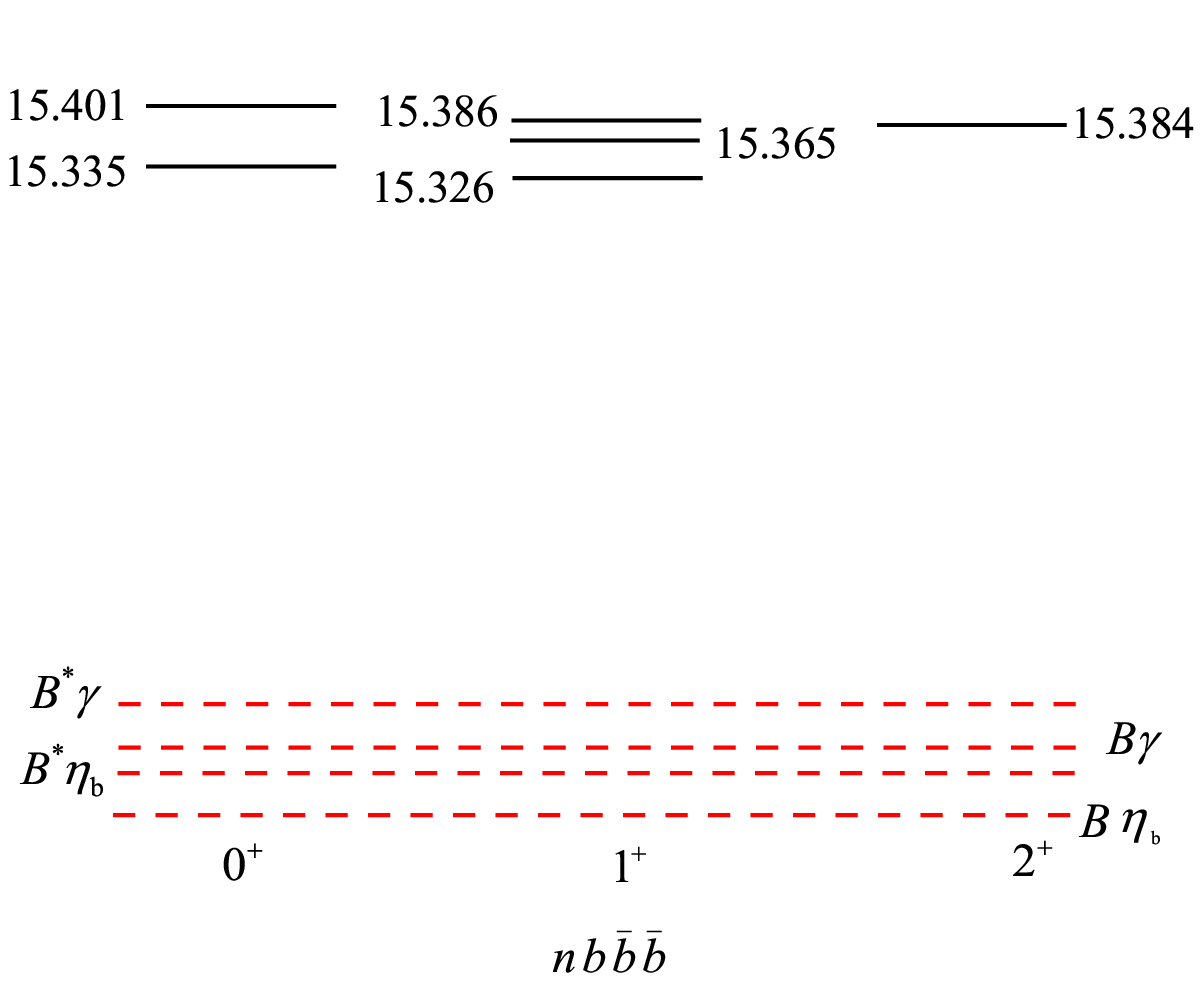} \includegraphics[width=0.5%
\textwidth]{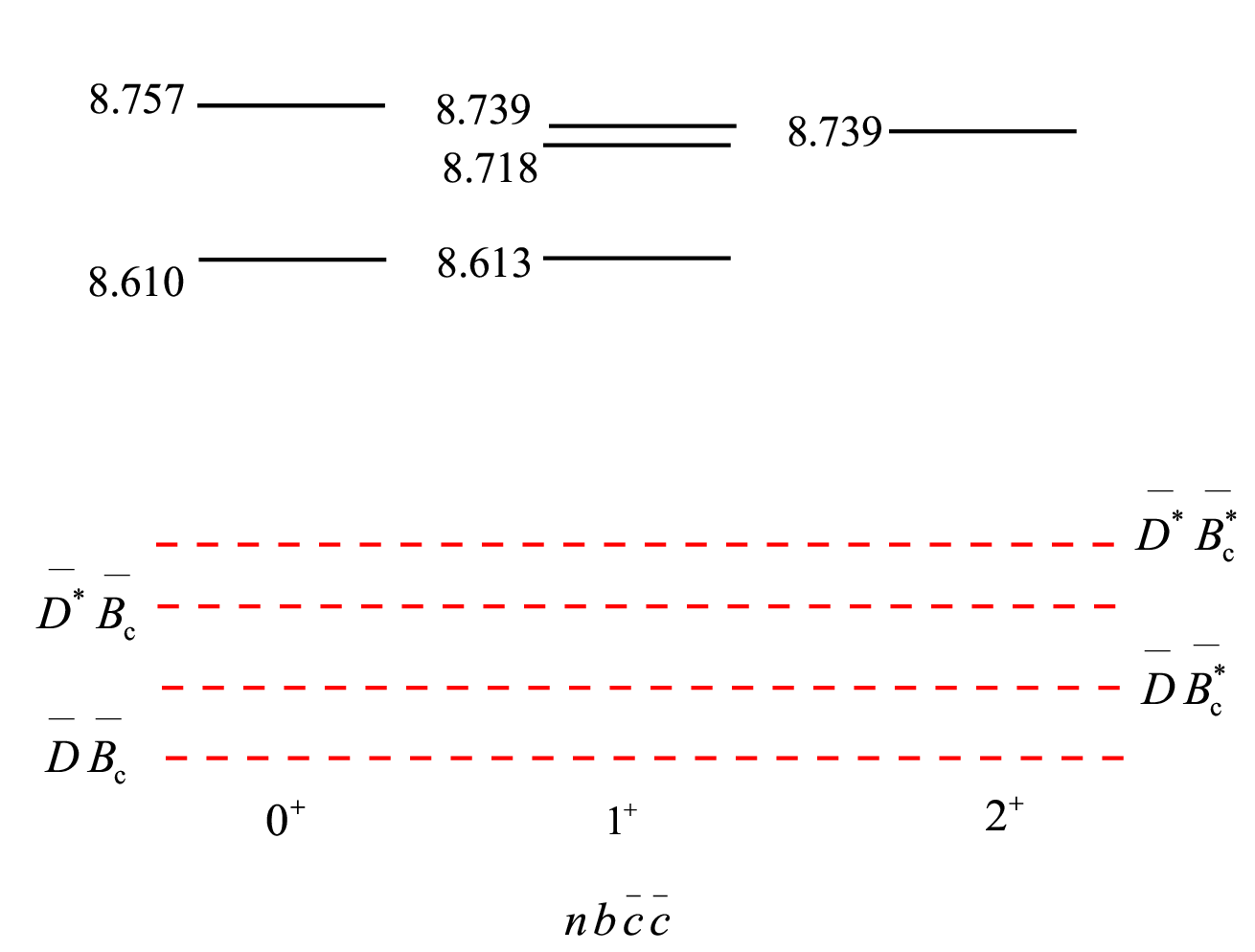}
\end{center}
\caption{Mass spectra(GeV) of the $nb\bar{b}\bar{b}$ and $nb\bar{c}\bar{c}$
tetraquark states(sold line), with meson-meson thresholds(GeV) also plotted
in dotted lines. }
\label{fig:1}
\end{figure}

First of all, we write the mass formula for the triply heavy tetraquarks $nb%
\bar{Q}\bar{Q}$ and $nc\bar{Q}\bar{Q}$ with $J^{P}=2^{+},1^{+},0^{+}$, which
are obtained via diagonalization of the CMI for $J^{P}=1^{+}\ $and $0^{+}$
multiplets(mixed states). Then, one can minimize the diagonalized mass (\ref%
{MBm}) using the variantional method to find the respective optimal bag
radii $R$ of the tetraquark and to solve $x_{i}$ depending on $R$ via the
transcendental equation(\ref{transc}) interactively. This leads to the
tetraquark masses (\ref{MBm}) for each of the triply heavy tetraquark
states, as shown in FIG. 1 and FIG. 2. Further, one can use Eqs. (\ref{rEsum}%
) and (\ref{musum}) to compute their charge radii and magnetic moments, and
use Eq. (\ref{decy}) to estimate the relative decay widths of the tetraquark
for the two-meson channel depicted in FIG. \ref{fig:1}.
\setlength{\abovecaptionskip}{0.2cm} \vspace{0.2cm}

\begin{figure}[tph]
\setlength{\abovecaptionskip}{0.3cm}
\par
\begin{center}
\includegraphics[width=0.5\textwidth]{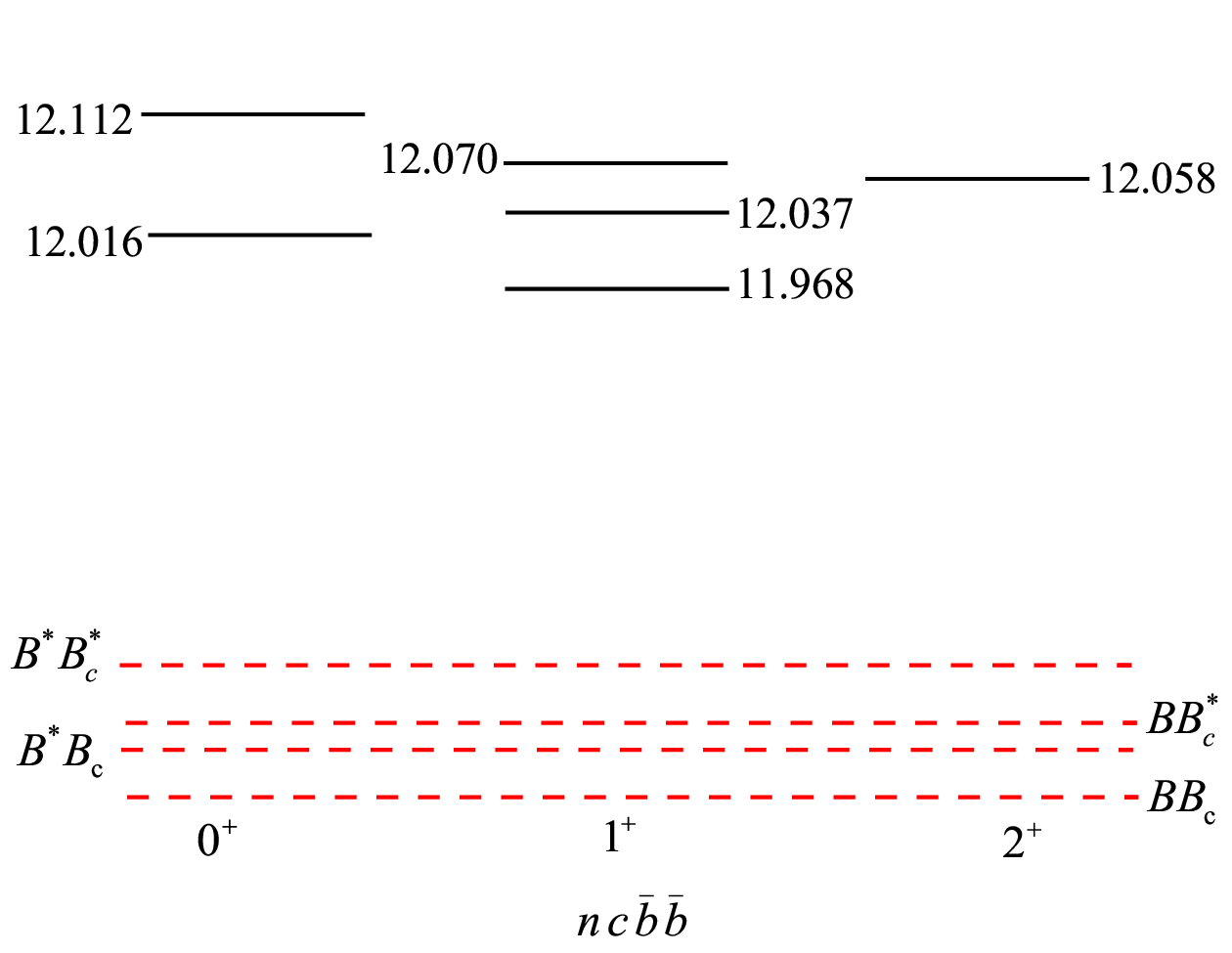} \includegraphics[width=0.5%
\textwidth]{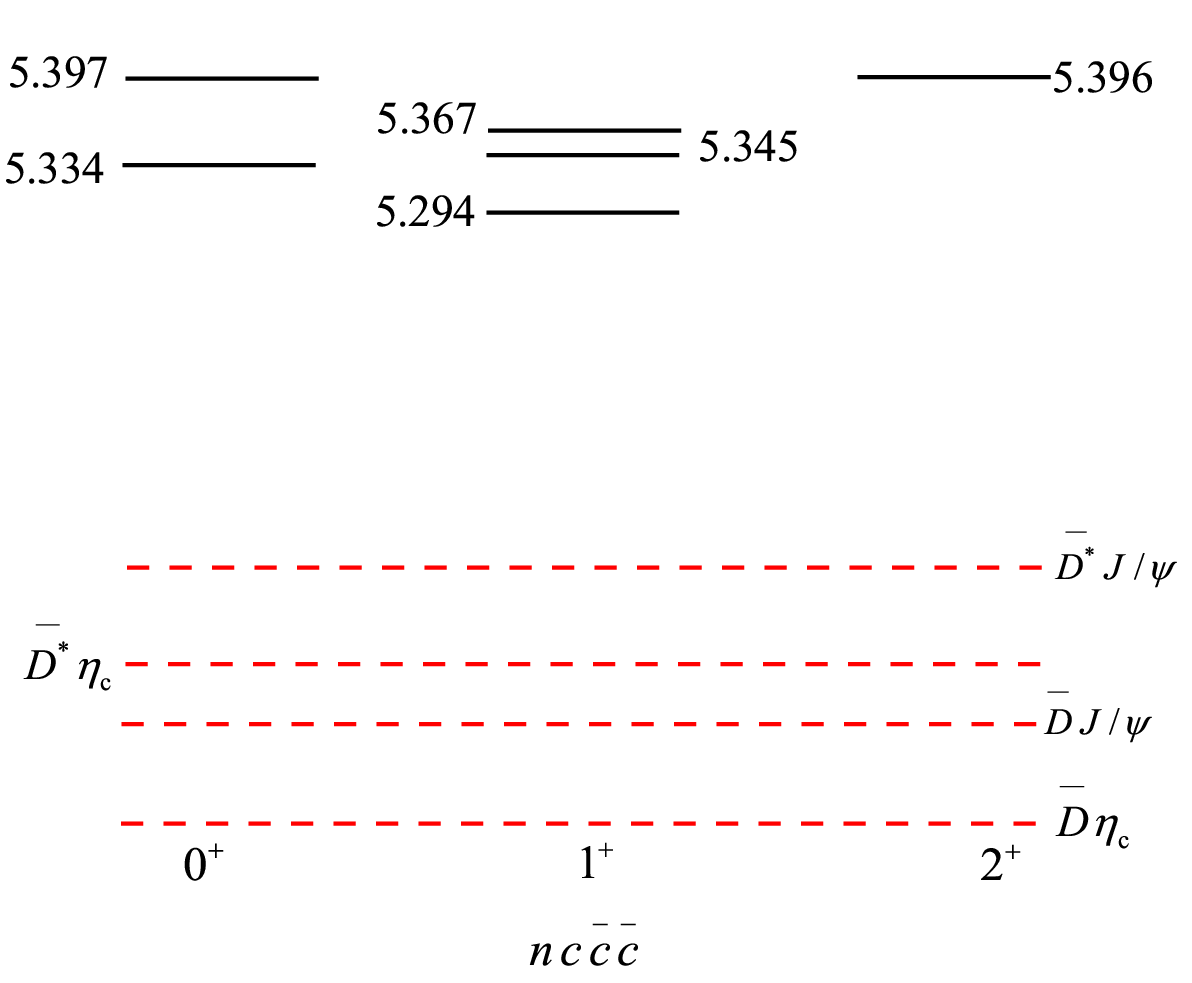}
\end{center}
\caption{Mass spectra(GeV) of the $nc\bar{b}\bar{b}$ and $nc\bar{c}\bar{c}$
tetraquark states in sold line, with meson-meson thresholds (GeV) plotted in
dotted lines. }
\label{fig:2}
\end{figure}
\setlength{\abovecaptionskip}{0.2cm} \renewcommand{\tabcolsep}{0.55cm} %
\renewcommand{\arraystretch}{1.5}
\begin{table*}[tbh]
\caption{Computed masses(in GeV), magnetic moments(in $\protect\mu _{N}$) and
charge radii (in $fm$) of triply heavy tetraquarks $nb\bar{Q}\bar{Q}$. Bag radii $R_{0}$ is in GeV$^{-1}$. The first and second values charge radii (magnetic moment) correspond to n=u and n=d, respectively. }
\label{tab:3heavytetraquark1}%
\begin{tabular}{cccccccc}
\hline\hline
\textrm{State} & $J^{P}$ & \textrm{Eigenvector} & $R_{0}$ & $M_{bag}$ & $%
r_{E}$ (fm) & $\mu_{bag}$ & \cite{Xin:2022} \\ \hline
${(nb\bar{b}\bar{b})}$ & ${2}^{+}$ & 1.00 & 4.14 & 15.384 & 0.55, 0.23 &
1.56, -0.65 & 14.872 \\
& ${1}^{+}$ & (-0.65,-0.19, 0.74) & 4.12 & 15.386 & 0.55, 0.23 & 0.82, -0.47
& 14.852 \\
&  & (0.64, -0.67, 0.39) & 4.07 & 15.365 & 0.54, 0.22 & 1.52, -0.56 & 14.852
\\
&  & (0.42, 0.72, 0.55) & 3.96 & 15.326 & 0.53,0.22 & -0.02,0.06 & 14.712 \\
& ${0}^{+}$ & (0.58, 0.81) & 4.19 & 15.401 & 0.56, 0.23 & - & 14.706 \\
&  & (-0.81, 0.58) & 3.99 & 15.335 & 0.53,0.22 & - & 14.851 \\
${(nb\bar{c}\bar{c})}$ & ${2}^{+}$ & 1.00 & 4.77 & 8.739 & 0.40, 0.79 &
0.63, -1.92 & 8.478 \\
& ${1}^{+}$ & (-0.63, -0.35, 0.70) & 4.74 & 8.739 & 0.40, 0.80 & 0.23, -0.71
& 8.454 \\
&  & (0.60, -0.79, 0.13) & 4.67 & 8.718 & 0.39, 0.78 & 0.70, -1.47 & 8.430
\\
&  & (0.50, 0.50, 0.70 ) & 4.52 & 8.613 & 0.38, 0.75 & -0.05, -0.66 & 8.218
\\
& ${0}^{+}$ & (0.69, 0.73) & 4.84 & 8.757 & 0.81, 0.81 & - & 8.445 \\
&  & (-0.73, 0.68 ) & 4.54 & 8.610 & 0.76, 0.76 & - & 8.199 \\ \hline\hline
${(nc\bar{b}\bar{b})}$ & ${2}^{+}$ & 1.00 & 4.50 & 12.058 & 0.78,0.44 &
2.26, -0.14 & 11.652 \\
& ${1}^{+}$ & (-0.36, -0.17, 0.92) & 4.40 & 12.070 & 0.76, 0.43 & 1.80,-0.15
& 11.659 \\
&  & (0.93, -0.19,0.33) & 4.40 & 12.037 & 0.76, 0.43 & 1.45, -0.40 & 11.625
\\
&  & (0.12, 0.97,0.23) & 4.25 & 11.968 & 0.74, 0.41 & 0.11,0.35 & 11.582 \\
& ${0}^{+}$ & (0.38, 0.93) & 4.55 & 12.112 & 0.79, 0.44 & - & 11.695 \\
&  & (-0.92, 0.38) & 4.36 & 12.016 & 0.76, 0.42 & - & 11.582 \\
${(nc\bar{c}\bar{c})}$ & ${2}^{+}$ & 1.00 & 5.01 & 5.396 & 0.37, 0.62 &
1.29, -1.39 & 5.198 \\
& ${1}^{+}$ & (-0.87, -0.05, 0.49) & 4.88 & 5.367 & 0.36, 0.61 & 0.92, -0.52
& 5.154 \\
&  & (0.47, -0.37, 0.80) & 4.86 & 5.345 & 0.35, 0.60 & 0.92, -0.52 & 5.136
\\
&  & (0.14, 0.93, 0.35) & 4.76 & 5.294 & 0.35, 0.59 & 0.89, -0.51 & 4.968 \\
& ${0}^{+}$ & (0.36, 0.83) & 5.01 & 5.397 & 0.37, 0.62 & - & 4.937 \\
&  & (-0.83, 0.55) & 4.87 & 5.334 & 0.36, 0.60 & - & 5.185 \\ \hline\hline
\end{tabular}%
\end{table*}

In Table. \ref{tab:3heavytetraquark1}, we list the obtained mass(in GeV),
magnetic moments(in $\mu _{N}$) and charge radii (in $f$m) of the triply
heavy tetraquarks $nb\bar{Q}\bar{Q}$ and $nc\bar{Q}\bar{Q}$, which contain
four states with given $J^{P}$: $nb\bar{b}\bar{b}$, $nb\bar{c}\bar{c}$, $nc%
\bar{b}\bar{b}$, and $nc\bar{c}\bar{c}$. For the tetraquark $nb\bar{b}\bar{b}
$, the $J^{P}=0^{+}$ state has mass of $15.335$ GeV for the mixed configuration dominated relatively by $%
\bar{3}_{c}\otimes {3}_{c}$ and has the mass of $15.401$ GeV for the mixed configuration dominated relatively by $%
6_{c}\otimes \bar{6}_{c}$. The later (in $6_{c}\otimes \bar{6}_{c}$) are
heavier than the former(in $\bar{3}_{c}\otimes {3}_{c}$) about $0.066$ GeV.
As depicted in FIG. \ref{fig:1}, this state may decay to the final two-meson
states of the $B^{\ast }\Upsilon $, $B\Upsilon $, $B^{\ast }\eta _{b}$ or $%
B\eta _{b}$. Our computation via Eq. (\ref{decy}) shows that the tetraquark $%
T(nb\bar{b}\bar{b},15.401,J^{P}=0^{+})$ can predominantly decay to $B^{\ast
}\Upsilon $, with the relative decay widths of $\Gamma _{B^{\ast }\Upsilon }$
: $\Gamma _{B\eta _{b}}$ $\sim $ $1.15:0.004$. In addition, it turns out
that tetraquark $T(nb\bar{b}\bar{b},15.386,J^{P}=1^{+})$ can predominantly
decay into $B^{\ast }\Upsilon $ or $B^{\ast }\eta _{b}$, with relative decay
widths of $\Gamma _{B^{\ast }\Upsilon }$ : $\Gamma _{B^{\ast }\eta _{b}}$ : $%
\Gamma _{B\Upsilon }$ $\sim $ $0.52:0.56:0.001$.

For the tetraquark state $nb\bar{c}\bar{c}$, the $J^{P}=0^{+}$ state has mass of $8.757$ GeV in the mixed state with
configuration $6_{c}\otimes \bar{6}_{c}$ dominated relatively and has mass of 
$8.610$ GeV with $\bar{3}_{c}\otimes 3_{c}$ dominated relatively, with the
later lighter than the former about $0.147$ GeV. As indicated in FIG. %
\ref{fig:1}, this state may decay to $\bar{D}^{\ast }\bar{B}_{c}^{\ast }$, $%
\bar{D}^{\ast }\bar{B}_{c}$, $\bar{D}\bar{B}_{c}^{\ast }$ or $\bar{D}\bar{B}%
_{c}$. The tetraquark $T(nb\bar{c}\bar{c},8.757,J^{P}=0^{+})$ can decay
dominantly to the channel $\bar{D}^{\ast }\bar{B}_{c}^{\ast }$, while the
tetraquark $T(nb\bar{c}\bar{c},8.739,J^{P}=1^{+})$ can predominantly decay
to $\bar{D}^{\ast }\bar{B}_{c}^{\ast }$ or $\bar{D}^{\ast }\bar{B}_{c}$. The
relative widths of them are $\Gamma _{\bar{D}^{\ast }\bar{B}_{c}^{\ast }}$ :
$\Gamma _{\bar{D}\bar{B}_{c}}$ $\sim $ $0.59:0.003$ and $\Gamma _{\bar{D}^{\ast }\bar{B}_{c}^{\ast }}$ :
$\Gamma _{\bar{D}^{\ast }\bar{B}_{c}}$ :
$\Gamma _{\bar{D}\bar{B}_{c}^{\ast }}$ $\sim $ $0.34:0.27:0.006$.

For the tetraquark $nc\bar{b}\bar{b}$, the $J^{P}=0^{+}$ state has mass of $12.016$ GeV in the mixed state with
configuration $\bar{3}_{c}\otimes 3_{c}$ dominated relatively and has mass of 
$12.112$ GeV with $6_{c}\otimes \bar6_{c}$ dominated relatively, with the
later heavier than the former about $0.096$ GeV. As shown in FIG. \ref{fig:2},
this state may decay to $B^{\ast }{B}_{c}^{\ast }$, $B{B}_{c}^{\ast }$, $%
B^{\ast }{B}_{c}$ or $B{B}_{c}$. The dominant decay chnnel of the tetraquark
$T(nc\bar{b}\bar{b},12.112,J^{P}=0^{+})$ is $B^{\ast }{B}_{c}^{\ast }$,
while the tetraquark $T(nc\bar{b}\bar{b},12.070,J^{P}=1^{+})$ can
predominantly decay to $B^{\ast }{B}_{c}^{\ast }$ or $B^{\ast }$${B}_{c}$.
The relative decay widths $\Gamma _{B^{\ast }{B}_{c}^{\ast }}$ : $\Gamma _{B{%
B}_{c}}$ $\sim $ $0.95:0.063$ and $\Gamma _{B^{\ast }{B}_{c}^{\ast }}
$ : $\Gamma _{B^{\ast }{B}_{c}}$ : $\Gamma _{B{B}_{c}^{\ast }}$ $\sim $ $%
0.56:0.37:0.06$.

For the tetraquark $nc\bar{c}\bar{c}$, the $J^{P}=0^{+}$ state has mass of $5.334$ GeV in the mixed state with
configuration $\bar3_{c}\otimes {3}_{c}$ dominated relatively and has mass of 
$5.397$ GeV with ${6}_{c}\otimes \bar6_{c}$ dominated relatively, with the
later heavier than the former about $0.063$ GeV. As depicted in FIG. \ref{fig:2}, the $%
0^{+}$ state can decay to $\bar{D}^{\ast }J/\psi $, $\bar{D}^{\ast }\eta
_{c} $, $\bar{D}J/\psi $ or $\bar{D}\eta _{c}$. The dominant channel of the
tetraquark $T(nc\bar{c}\bar{c},5.397,J^{P}=0^{+})$ is the $\bar{D}^{\ast
}J/\psi $, while the tetraquark $T(nc\bar{c}\bar{c},5.367,J^{P}=1^{+})$ can
predominantly decay to $\bar{D}^{\ast }J/\psi $ or $\bar{D}^{\ast }\eta _{c}$%
, with the relative width $\Gamma _{\bar{D}^{\ast }{J/\psi }}$ : $\Gamma _{%
\bar{D}\eta _{c}}$ $\sim $ $0.36:0.04$. The decay width ratio is $\Gamma _{%
\bar{D}^{\ast }{J/\psi }}$ : $\Gamma _{\bar{D}^{\ast }{\eta _{c}}}$ : $%
\Gamma _{\bar{D}{J/\psi }}$ $\sim $ $0.12:0.14:0.07$.

In Table. \ref{tab:3heavytetraquark1}, we also compare our calculated
tetraquark masses to that in Ref. \cite{Xin:2022}, with mass difference
about $0.2-0.5$ GeV.

\subsection{Triply heavy tetraquarks $sb\bar{Q}\bar{Q}$ and $sc\bar{Q}\bar{Q}
$.}

Due to chromomagnetic interaction, the triply heavy tetraquarks $sb\bar{Q}%
\bar{Q}$ and $sc\bar{Q}\bar{Q}$ can mix their spin-color bases to form mixed
states, a linear combination of spin-color basis functions ($\phi \chi $). A
$J^{P}=2^{+}$ state of tetraquark consists of a singlet $\phi _{2}\chi _{1}$
while the $J^{P}=1^{+}$ state consist of the mixed states $c_{1}\phi
_{2}\chi _{2}+c_{2}\phi _{2}\chi _{5}+c_{3}\phi _{1}\chi _{4}$ in the
subspace ($\phi _{2}\chi _{2},\phi _{2}\chi _{5},\phi _{1}\chi _{4}$); the $%
J^{P}=0^{+}$ state is composed of the mixed state $c_{1}\phi _{2}\chi
_{3}+c_{2}\phi _{1}\chi _{6}$ in the subspace ($\phi _{2}\chi _{3},c_{2}\phi
_{1}\chi _{6}$). Similarly, we employ MIT bag mode to compute the masses, bag
radii, magnetic moments, and charge radii of the triply heavy tetraquarks $%
sb\bar{Q}\bar{Q}$ and $sc\bar{Q}\bar{Q}$ with the spin-color wavefunctions.
The results are shown in Table. \ref{tab:3heavytetraquark2}.

\renewcommand{\tabcolsep}{0.7cm} \renewcommand{\arraystretch}{1.5}
\begin{table*}[tbh]
\caption{Computed masses(in GeV), magnetic moments and charge radii of
triply heavy tetraquarks $sb\bar{Q}\bar{Q}$ and $sc\bar{Q}\bar{Q}$. Bag
radii  $R_{0}$ is in GeV$^{-1}$. }
\label{tab:3heavytetraquark2}%
\begin{tabular}{cccccccc}
\hline\hline
\textrm{State} & $J^{P}$ & \textrm{Eigenvector} & $R_{0}$ & $M_{bag}$ & $%
r_{E}$ (fm) & $\mu_{bag}$ & \cite{Xin:2022} \\ \hline
${(sb\bar{b}\bar{b})}$ & ${2}^{+}$ & 1.00 & 4.22 & 15.476 & 0.19 & -0.52 &
14.957 \\
& ${1}^{+}$ & (-0.63, -0.21, 0.75) & 4.07 & 15.476 & 0.19 & -0.38 & 14.944
\\
&  & (0.66, -0.66, 0.37) & 4.02 & 15.458 & 0.19 & -0.43 & 14.929 \\
&  & (0.41, 0.73, 0.55) & 3.91 & 15.424 & 0.18 & -0.06 & 14.805 \\
& ${0}^{+}$ & (0.58, 0.81) & 4.26 & 15.492 & 0.20 & - & 14.936 \\
&  & (-0.81, 0.58) & 4.09 & 15.433 & 0.19 & - & 14.793 \\
${(sb\bar{c}\bar{c})}$ & ${2}^{+}$ & 1.00 & 4.82 & 8.840 & 0.79 & -1.74 &
8.569 \\
& ${1}^{+}$ & (-0.83, -0.07, 0.55) & 4.65 & 8.823 & 0.76 & -0.78 & 8.553 \\
&  & (0.52, -0.48, 0.71) & 4.64 & 8.794 & 0.76 & -0.77 & 8.522 \\
&  & (0.22, 0.87, 0.44) & 4.53 & 8.736 & 0.74 & -0.77 & 8.325 \\
& ${0}^{+}$ & (0.73, 0.68) & 4.84 & 8.839 & 0.79 & - & 8.301 \\
&  & (-0.69, 0.72) & 4.70 & 8.783 & 0.77 & - & 8.539 \\ \hline\hline
${(sc\bar{b}\bar{b})}$ & ${2}^{+}$ & 1.00 & 4.55 & 12.158 & 0.46 & 2.26 &
11.652 \\
& ${1}^{+}$ & (-0.37,-0.19,0.91) & 4.36 & 12.162 & 0.44 & 0.01 & 11.659 \\
&  & (0.92,-0.19,0.34) & 4.35 & 12.138 & 0.44 & -0.27 & 11.625 \\
&  & (0.11,0.96,0.25) & 4.21 & 12.078 & 0.43 & 0.35 & 11.582 \\
& ${0}^{+}$ & (0.40,0.91) & 4.60 & 12.200 & 0.47 & - & 11.582 \\
&  & (-0.91,0.41) & 4.43 & 12.119 & 0.45 & - & 11.695 \\
${(sc\bar{c}\bar{c})}$ & ${2}^{+}$ & 1.00 & 5.05 & 5.504 & 0.60 & -1.19 &
5.303 \\
& ${1}^{+}$ & (-0.43, 0.40, 0.81) & 4.90 & 5.499 & 0.59 & -0.10 & 5.254 \\
&  & (0.85, -0.49, 0.20) & 4.83 & 5.464 & 0.58 & -1.35 & 5.240 \\
&  & (0.31, 0.77, 0.55) & 4.68 & 5.391 & 0.56 & -0.30 & 5.069 \\
& ${0}^{+}$ & (0.58,0.81) & 5.13 & 5.540 & 0.61 & - & 5.040 \\
&  & (-0.81, 0.58) & 4.80 & 5.391 & 0.58 & - & 5.291 \\ \hline\hline
\end{tabular}%
\end{table*}
\vspace{0.2cm}

\begin{figure}[th]
\setlength{\abovecaptionskip}{0.3cm}
\par
\begin{center}
\includegraphics[width=0.5\textwidth]{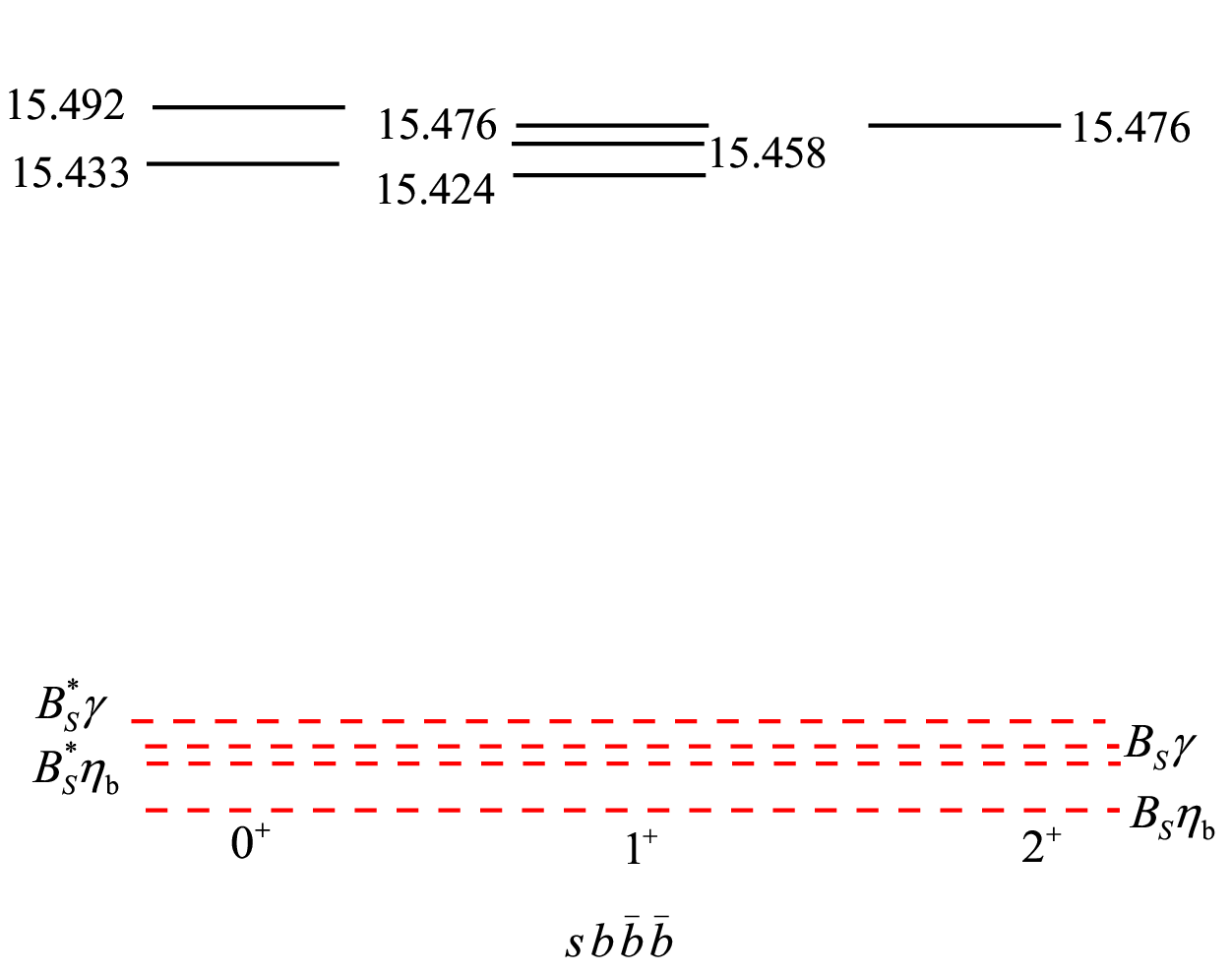} \includegraphics[width=0.5%
\textwidth]{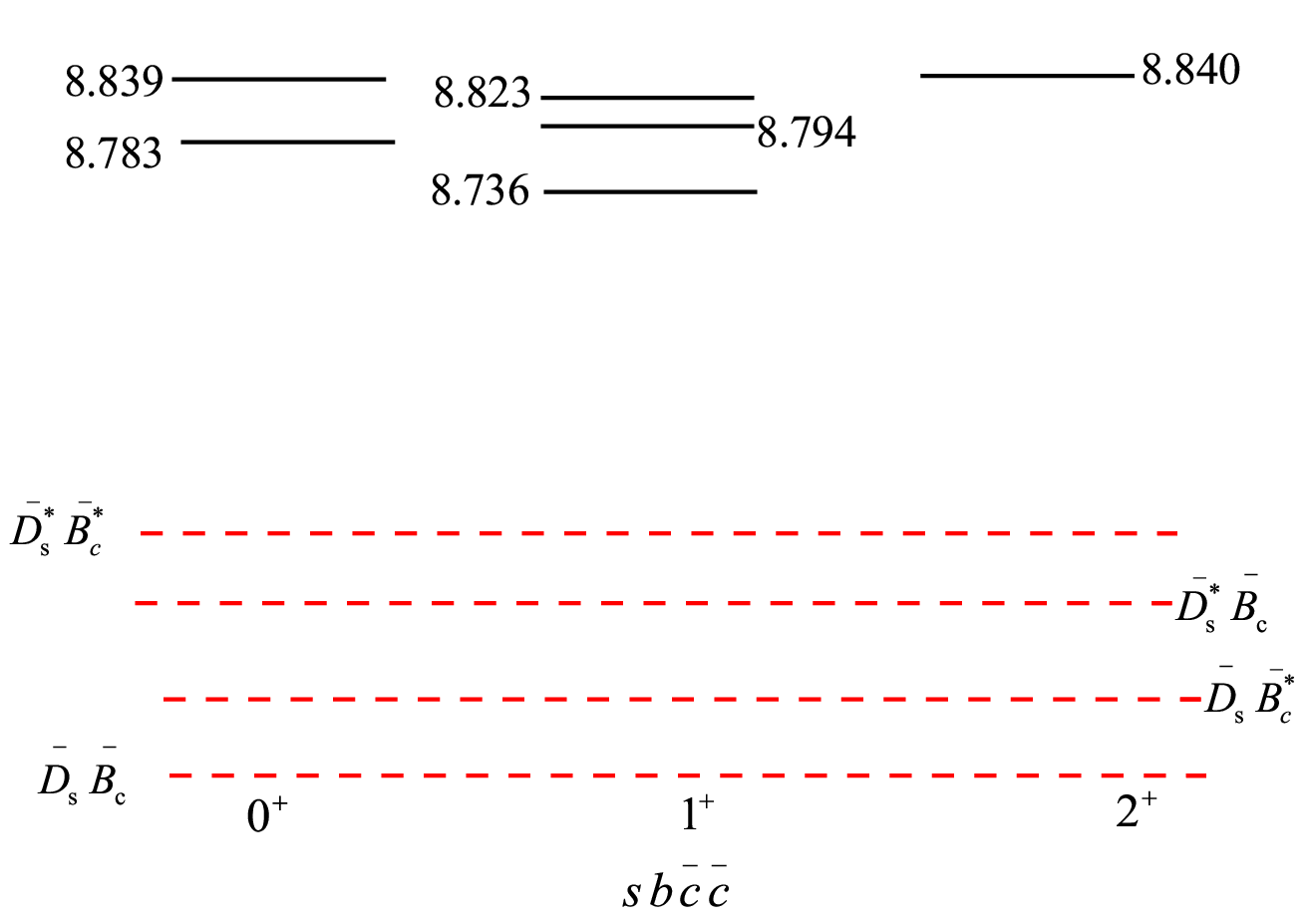}
\end{center}
\caption{Mass spectra(GeV) of the tetraquarks $sb\bar{b}\bar{b}$ and $sb\bar{%
c}\bar{c}$ plotted in sold line. The meson-meson thresholds (GeV) are also
shown in dotted lines.}
\label{fig:3}
\end{figure}
\setlength{\abovecaptionskip}{0.2cm}
\begin{figure}[th]
\setlength{\abovecaptionskip}{0.3cm}
\par
\begin{center}
\includegraphics[width=0.5\textwidth]{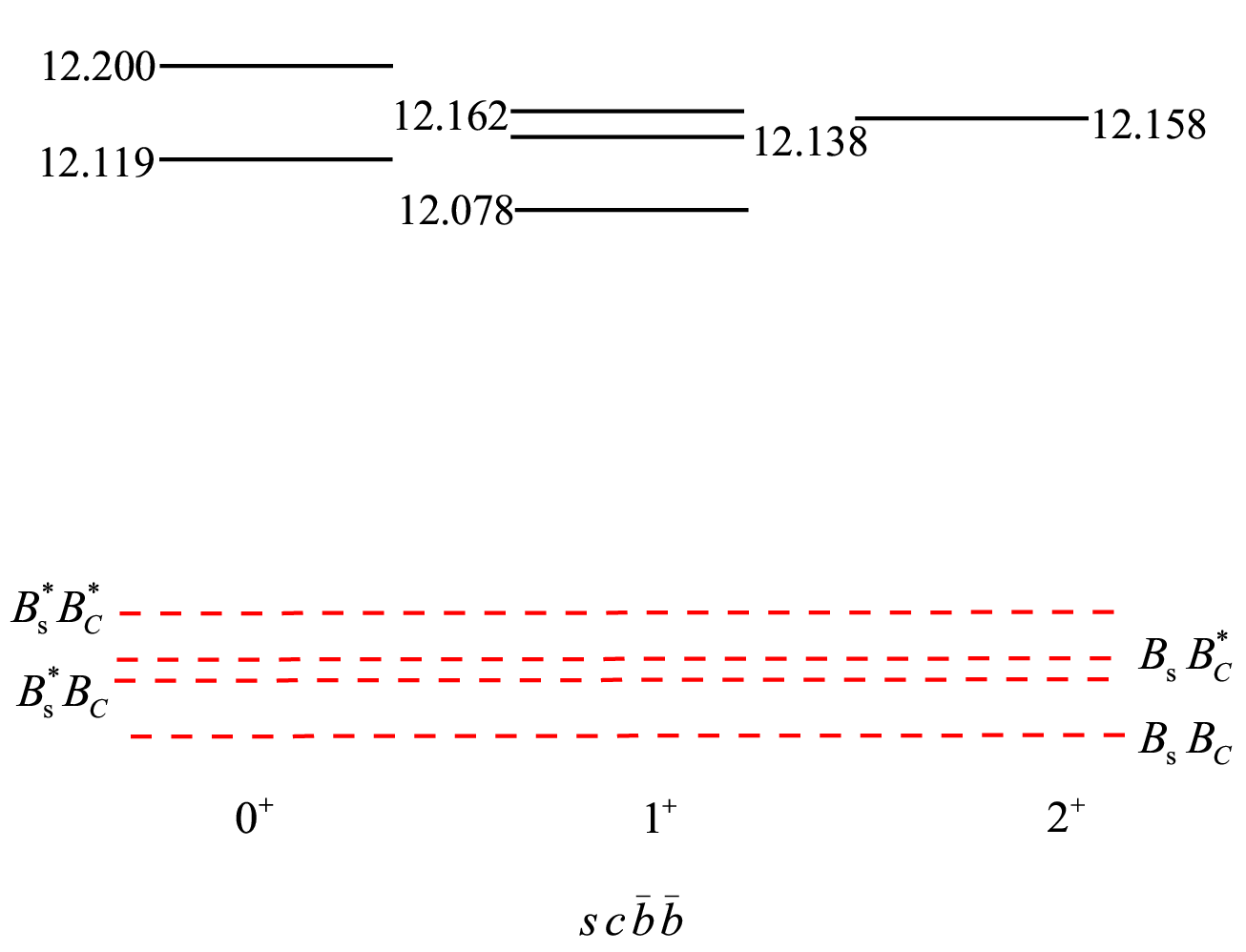} \includegraphics[width=0.5%
\textwidth]{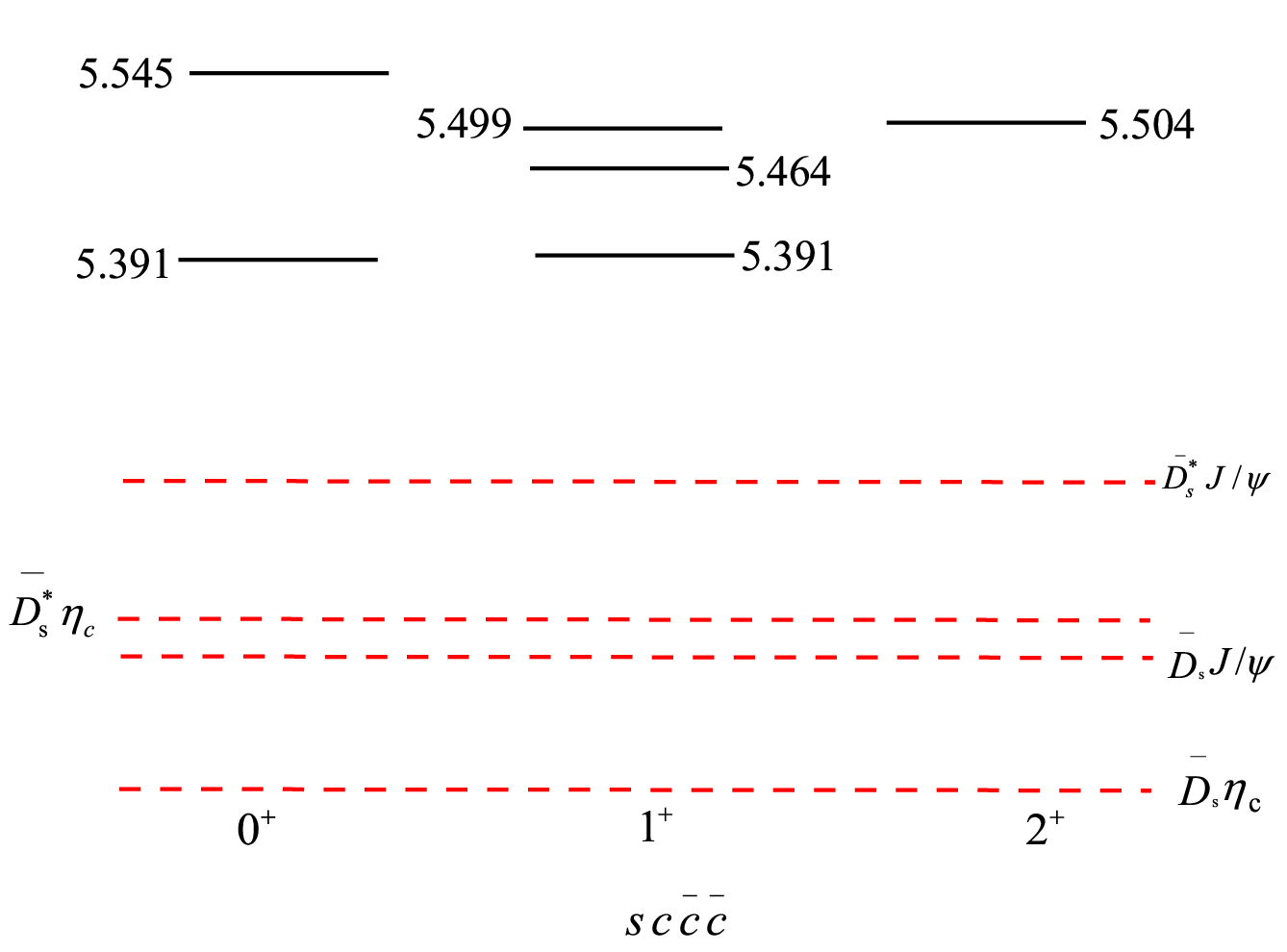}
\end{center}
\caption{Mass spectra(GeV) of the tetraquarks $sc\bar{b}\bar{b}$ and $sc\bar{%
c}\bar{c}$ plotted in sold line. The meson-meson thresholds (GeV) are shown
in dotted lines. }
\label{fig:4}
\end{figure}
\setlength{\abovecaptionskip}{0.2cm}

The triply heavy tetraquarks $sb\bar{Q}\bar{Q}$ and $sc\bar{Q}\bar{Q}$
consitst in flavor context of the four states: the tetraquark $sb\bar{b}\bar{%
b}$, the $sb\bar{c}\bar{c}$, the $sc\bar{b}\bar{b}$, and the $sc\bar{c}\bar{c%
}$. In the case of the $sb\bar{b}\bar{b}$, the $J^{P}=0^{+}$ state has mass of $15.433$ GeV in the mixed state with
configuration $\bar3_{c}\otimes {3}_{c}$ dominated relatively and has mass of 
$15.492$ GeV with ${6}_{c}\otimes\bar 6_{c}$ dominated relatively, with the
later heavier than the former about $0.059$ GeV. This state can decay into the final two-meson states of the $%
B_{s}^{\ast }\Upsilon $, $B_{s}\Upsilon $, $B_{s}^{\ast }\eta _{b}$ or $%
B_{s}\eta _{b}$, as shown in FIG. \ref{fig:3}. We find that relative decay
width between the channals $B_{s}^{\ast }\Upsilon $ and $B_{s}\eta _{b}$ is $%
\Gamma _{B_{s}^{\ast }\Upsilon }$ : $\Gamma _{B_{s}\eta _{b}}$ $\sim $ $%
0.07:0.94$. The state $T(sb\bar{b}\bar{b},15.492,0^{+})$ can dominantly
decay into $B_{s}\eta _{b}$ whereas the state $T(sb\bar{b}\bar{b}%
,15.476,1^{+})$ can primarily decay into $B_{s}^{\ast }\Upsilon $ or $%
B_{s}^{\ast }\eta _{b}$, with computed relative decay width $\Gamma
_{B_{s}^{\ast }\Upsilon }$ : $\Gamma _{B_{s}^{\ast }\eta _{b}}$ : $\Gamma
_{B_{s}\Upsilon }$ $\sim $ $0.53:0.55:0.001$.

Regarding the tetraquark $sb\bar{c}\bar{c}$,  $J^{P}=0^{+}$ state has mass of $8.783$ GeV in the mixed state with
configuration ${6}_{c}\otimes\bar 6_{c}$ dominated relatively and has mass of 
$8.839$ GeV with  $\bar3_{c}\otimes {3}_{c}$ dominated relatively, with the
later heavier than the former about $0.056$ GeV. As illustrated in FIG. \ref{fig:3}, the tetraquark $sb%
\bar{c}\bar{c}$ can potentially decay to $\bar{D}_{s}^{\ast }\bar{B}%
_{c}^{\ast }$, $\bar{D}_{s}^{\ast }\bar{B}_{c}$, $\bar{D}_{s}\bar{B}%
_{c}^{\ast }$ or $\bar{D}_{s}\bar{B}_{c}$, with the relative decay widths
about $\Gamma _{\bar{D}_{s}^{\ast }\bar{B}_{c}^{\ast }}$ : $\Gamma _{\bar{D}%
_{s}\bar{B}_{c}}$ $\sim $ $0.59:0.001$. The tetraquark $T(sb\bar{c}\bar{c}%
,8.839,0^{+})$ can decay into $\bar{D}_{s}^{\ast }\bar{B}_{c}^{\ast }$
dominantly, while $T(sb\bar{c}\bar{c},8.823,1^{+})$ can primarily decay into
$\bar{D}_{s}^{\ast }\bar{B}_{c}^{\ast }$ or $\bar{D}_{s}^{\ast }\bar{B}_{c}$%
, with relative decay width of $\Gamma _{\bar{D}_{s}^{\ast }\bar{B}%
_{c}^{\ast }}$ $:$ $\Gamma _{\bar{D}_{s}^{\ast }\bar{B}_{c}}$ $:$ $\Gamma _{%
\bar{D}_{s}\bar{B}_{c}^{\ast }}$ $\sim $ $0.14:0.24:0.009$.

For the tetraquark $sc\bar{b}\bar{b}$, the $0^{+}$ state has the mass of $%
12.119$ GeV for the dominated rep. $\bar{3}_{c}\otimes {3}_{c}$ configuration and the mass
of $12.200$ GeV for the dominated rep $6_{c}\otimes \bar{6}_{c}$ (heavier than the former about
$0.081$ GeV). As shown in FIG. \ref{fig:4}, the decay channels of this state
contain potentially $B_{s}^{\ast }B_{c}^{\ast }$, $B_{s}B_{c}^{\ast }$, $%
B_{s}^{\ast }B_{c}$, and $B_{s}B_{c}$. The state $T(sc\bar{b}\bar{b}%
,12.200,0^{+})$ can decay to $B_{s}^{\ast }B_{c}^{\ast }$, whereas $T(sc\bar{%
b}\bar{b},12.162,J^{P}=1^{+})$ to $B_{s}^{\ast }B_{c}^{\ast }$ or $%
B_{s}^{\ast }$$B_{c}$, with computed relative decay widths $\Gamma
_{B_{s}^{\ast }B_{c}^{\ast }}$:$\Gamma _{B_{s}B_{c}}$ $\sim $ $0.95:0.05$
and $\Gamma _{B_{s}^{\ast }B_{c}^{\ast }}$:$\Gamma _{B_{s}^{\ast }B_{c}}$:$%
\Gamma _{B_{s}B_{c}^{\ast }}$ $\sim $ $0.57:0.37:0.05$.

In the case of the tetraquark $sc\bar{c}\bar{c}$, the $0^{+}$ state has the
mass of $5.391$ GeV for the dominated rep. $\bar{3}_{c}\otimes {3}_{c}$ configuration and the
mass of $5.545$ GeV for the dominated rep. $6_{c}\otimes \bar{6}_{c}$(heavier than the former $%
0.154$ GeV). As depicted in FIG. \ref{fig:4}, the potential decay channels
of this state contan $\bar{D}_{s}^{\ast }J/\psi $, $\bar{D}_{s}^{\ast }\eta
_{c}$, $\bar{D}_{s}J/\psi $, and $\bar{D}_{s}\eta _{c}$. The state $T(sc\bar{%
c}\bar{c},5.540,0^{+})$ can decay to $\bar{D}_{s}^{\ast }J/\psi $
dominantly, while the state $T(sc\bar{c}\bar{c},5.499,1^{+})$ can primarily
decay to $\bar{D}_{s}^{\ast }J/\psi $ or $\bar{D}_{s}^{\ast }\eta _{c}$,
with the relative decay width $\Gamma
_{\bar{D}_{s}^{\ast }J/\psi }$:$\Gamma _{%
\bar{D}_{s}\eta _{c}}$ $\sim $ $0.51:0.002$ and  $\Gamma _{\bar{D}_{s}^{\ast }J/\psi }:\Gamma _{%
\bar{D}_{s}^{\ast }\eta _{c}}$ $:$ $\Gamma _{\bar{D}_{s}J/\psi }$ $\sim $ $%
0.34:0.16:0.001$.

\subsection{Triply heavy tetraquarks $nQ\bar{Q}\bar{Q\prime }$.}

\begin{figure}[th]
\setlength{\abovecaptionskip}{0.3cm}
\par
\begin{center}
\includegraphics[width=0.5\textwidth]{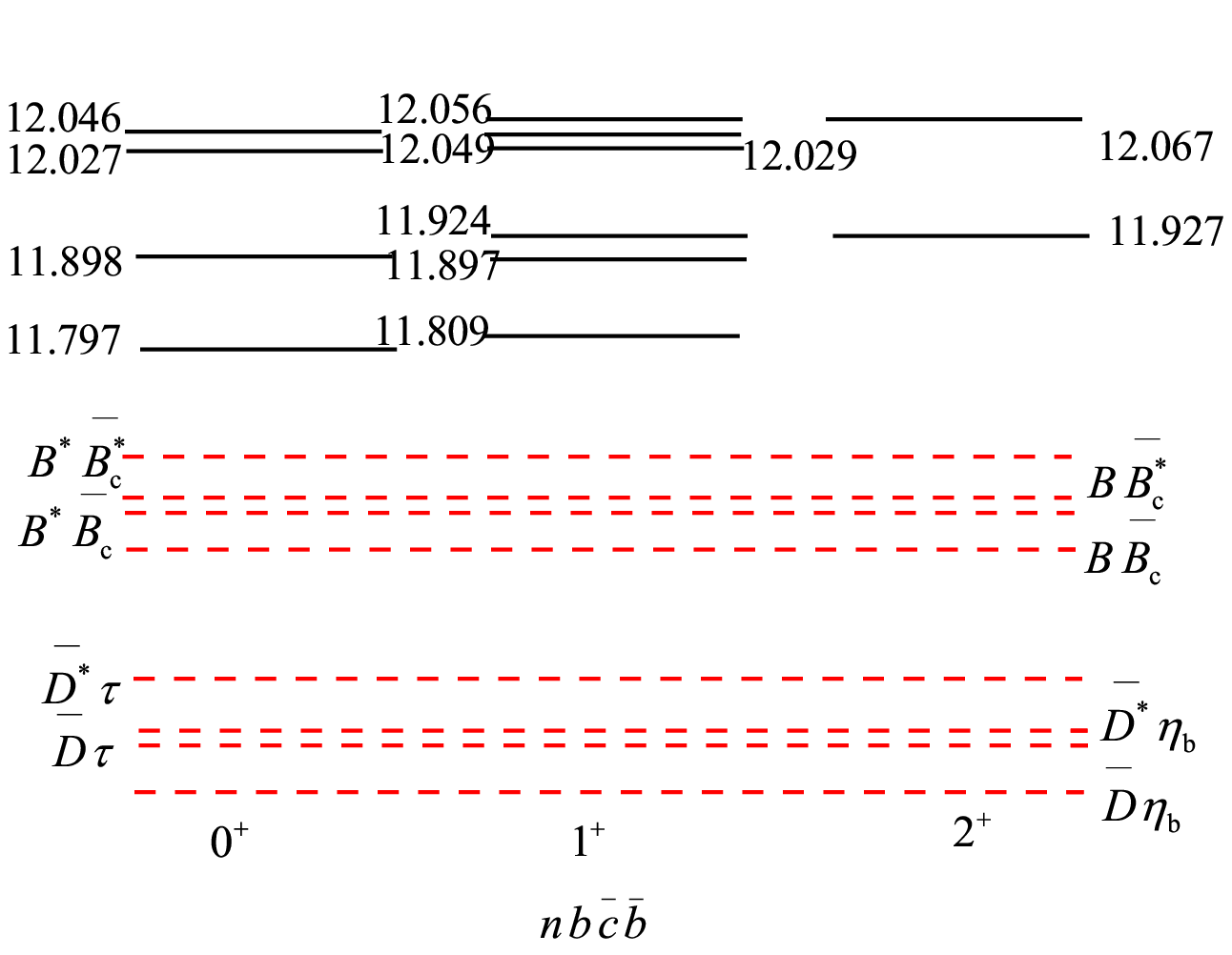} \includegraphics[width=0.5%
\textwidth]{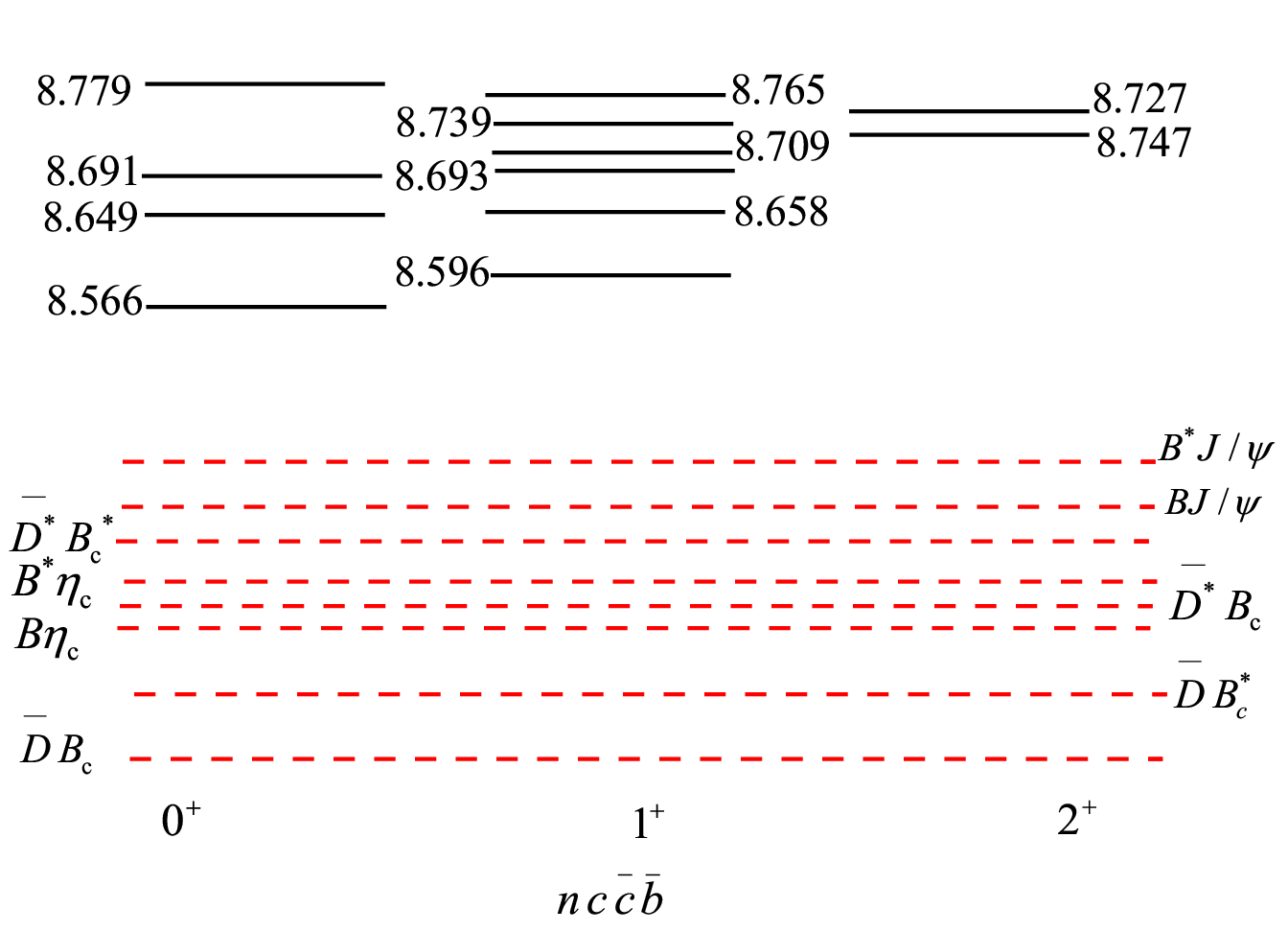}
\end{center}
\caption{Mass spectra(GeV, the sold lines) of the tetraquarks $nb\bar{c}\bar{%
b}$ and $nc\bar{c}\bar{b}$, with their two-meson thresholds(GeV) plotted as
dotted lines. }
\label{fig:5}
\end{figure}

This type of triply heavy tetraquarks consists of two classes of the
tetraquark states: $nb\bar{c}\bar{b}$ and $nc\bar{c}\bar{b}$. The state with
$J^{P}=2^{+}$ contains two configurations ($\phi _{2}\chi _{1}$, $\phi
_{1}\chi _{1}$) while the $1^{+}$ state, spaned by the basis states ($\phi
_{2}\chi _{2},\phi _{2}\chi _{4},\phi _{2}\chi _{5},\phi _{1}\chi _{2},\phi
_{1}\chi _{4},\phi _{1}\chi _{5}$), form a mixed state, $c_{1}\phi _{2}\chi
_{2}+c_{2}\phi _{2}\chi _{4}+c_{3}\phi _{2}\chi _{5}+c_{4}\phi _{1}\chi
_{2}+c_{5}\phi _{1}\chi _{4}+c_{6}\phi _{1}\chi _{5}$. Additionally, the $%
0^{+}$ state can be mixed state $c_{1}\phi _{2}\chi _{3}+c_{2}\phi _{2}\chi
_{6}+c_{3}\phi _{1}\chi _{3}+c_{4}\phi _{1}\chi _{6}$ in the space ($\phi
_{2}\chi _{3},\phi _{2}\chi _{6},\phi _{1}\chi _{3},\phi _{1}\chi _{6}$).
Treating the CMI as perturbation, one can calculate the mass and other
properties, as listed in Table \ref{tab:3heavytetraquark3}. Given the
computed masses (shown in in FIG. \ref{fig:5}) of the tetraquark $nc\bar{c}%
\bar{b}$, they exceed the threshold and may decay to $B^{\ast }J/\psi $, $%
BJ/\psi $, $\bar{D}^{\ast }\bar{B}_{c}^{\ast }$, $B^{\ast }\eta _{c}$, $\bar{%
D}^{\ast }\bar{B}_{c}$, $B\eta _{c}$, $\bar{D}^{\ast }\bar{B}_{c}^{\ast }$
or $\bar{D}\bar{B}_{c}$, as shown in FIG. \ref{fig:5}. Notably, the analysis
via eigenvector of the CMI matrix reveals that the state $T(nc\bar{c}\bar{b}%
,8.747,2^{+})$ couples srongly to $\bar{D}^{\ast }\bar{B}_{c}^{\ast }$,
while the $T(nc\bar{c}\bar{b},8.596,1^{+})$ couples significantly to $\bar{D}%
^{\ast }\bar{B}_{c}$. Furthermore, the state $T(nc\bar{c}\bar{b}%
,8.566,0^{+}) $ couples srongly to $\bar{D}\bar{B}_{c}$, a candidate of the
scattering states.

Regarding the tetraquark $nb\bar{c}\bar{b}$, the main decay channels contain
$B^{\ast }\bar{B}_{c}^{\ast }$, $B\bar{B}_{c}^{\ast }$, $B^{\ast }\bar{B}%
_{c} $, $B\bar{B}_{c}$, $\bar{D}^{\ast }\Upsilon $, $\bar{D}^{\ast }\eta _{b}
$, $\bar{D}\Upsilon $ and $\bar{D}\eta _{b}$, as depicted in FIG. \ref{fig:5}%
. Notably, analysis via comuted masses and eigenvectors (shown in Table V),
the state $T(nb\bar{c}\bar{b},11.927,2^{+})$ exhibit a pronounced coupling
to $\bar{D}^{\ast }\Upsilon $, while the $T(nb\bar{c}\bar{b},11.809,1^{+})$
states display a strong coupling to $\bar{D}\Upsilon $. Additionally, the $%
T(nb\bar{c}\bar{b},11.898,0^{+})$ states demonstrate a robust coupling to $%
B^{\ast }\bar{B}_{c}^{\ast }$, whereas the state $T(nb\bar{c}\bar{b}%
,11.797,0^{+})$ s exhibit a strong coupling to $\bar{D}\eta _{b}$,
potentially indicating scattering states.

\renewcommand{\tabcolsep}{0.6cm} \renewcommand{\arraystretch}{1.5}
\begin{table*}[tbh]
\caption{Computed masses (in GeV) triply heavy tetraquarks $nQ\bar{Q}\bar{%
Q\prime }$, with bag radii  $R_{0}$ in GeV$^{-1}$. The masses are compared
to other calculations cited. The charge radii are 0.32 fm (n=u) or 0.55 fm (n=d) for the tetraquark ${nb\bar{c}\bar{b}}$;They are 0.62 fm(n=u) or 0.26 fm(n=d) for the ${nc\bar{c}\bar{b}}$. The Bag radii  $R_{0}$ are 4.430 GeV$^{-1}$ for ${nb\bar{c}\bar{b}}$  and 4.700 GeV$^{-1}$ for ${nc\bar{c}\bar{b}}$. The first and second values magnetic moment correspond to n=u and n=d, respectively.}
\label{tab:3heavytetraquark3}%
\begin{tabular}{cccccccc}
\hline\hline
\textrm{State} & $J^{P}$ & \textrm{Eigenvector} & $M_{bag}$ & $\mu_{bag}$ &
\cite{Xin:2022} & \textrm{Scattering state} &  \\ \hline
${(nb\bar{c}\bar{b})}$ & ${2}^{+}$ & 1.00 & 11.927 & 1.09, -1.28 & 11.545 & $%
\bar{D}^{*}$$\Upsilon$ &  \\
&  & 1.00 & 12.067 & 1.09, -1.28 & 11.748 &  &  \\
& ${1}^{+}$ & (-0.15,0.17,-0.14,0.77,-0.37,0.44) & 11.809 & 0.63, -0.34 &
11.315 & $\bar{D}$$\Upsilon$ &  \\
&  & (-0.16,-0.11,-0.17,-0.34,-0.88,-0.22) & 11.897 & -0.02,-1.00 & 11.392 &
&  \\
&  & (-0.10,0.08,0.8,-0.52,-0.02,0.84) & 11.924 & -0.01, -0.59 & 11.447 &  &
\\
&  & (0.41,-0.21,0.83,0.11,-0.27,0.06) & 12.029 & 0.98, -0.74 & 11.698 &  &
\\
&  & (-0.63,0.57,0.48,-0.04,0.017,-0.20) & 12.049 & 2.00,-0.70 & 11.719 &  &
\\
&  & (0.61,0.76,-0.13,-0.11,-0.12,-0.05) & 12.056 & -0.28, -0.47 & 11.745 &
&  \\
& ${0}^{+}$ & (-0.16,-0.23,0.93,0.26) & 11.797 & - & 11.253 & $\bar{D}%
\eta_{b}$ &  \\
&  & (-0.32,0.10,-0.28,0.90) & 11.898 & - & 11.439 & $B^{*}\bar{B}_{c}^{*}$
&  \\
&  & (0.26,0.93,0.26,0.07) & 12.027 & - & 11.673 &  &  \\
&  & (0.90,-0.28,-0.013,0.35) & 12.046 & - & 11.749 &  &  \\
${(nc\bar{c}\bar{b})}$ & ${2}^{+}$ & 1.00 & 8.727 & 1.76, -0.75 & 8.335 & $%
\bar{D}^{*}$$\bar{B}_{c}^{*}$ &  \\
&  & 1.00 & 8.747 & 1.76, -0.75 & 8.469 &  &  \\
& ${1}^{+}$ & (0.27,-0.23,0.60,-0.56,0.39,-0.23) & 8.596 & 0.18, -0.14 &
8.190 &  &  \\
&  & (-0.14,0.43,0.69,0.42,0.11,0.35) & 8.658 & -0.11, -0.41 & 8.268 &  &
\\
&  & (-0.79,0.09,0.06,-0.55,-0.20,0.17) & 8.693 & 1.22, -0.54 & 8.313 &  &
\\
&  & (-0.20,-0.69,0.36,0.33,-0.47,-0.19) & 8.709 & 3.26, -0.64 & 8.399 &  &
\\
&  & (0.45,0.33,0.17,-0.29,-0.76,0.04) & 8.739 & 0.59, -0.17 & 8.436 &  &
\\
&  & (0.22,-0.42,-0.07,-0.13,0.03,0.87) & 8.765 & 0.15, -0.35 & 8.457 &  &
\\
& ${0}^{+}$ & (0.20,0.63,-0.74,-0.13) & 8.566 & - & 8.123 & $\bar{D}\bar{B}%
_{c}$ &  \\
&  & (0.80,-0.34,0.01,-0.50) & 8.649 & - & 8.281 &  &  \\
&  & (0.24,0.70,0.67,-0.07) & 8.691 & - & 8.388 &  &  \\
&  & (0.52,-0.04,-0.05,0.85) & 8.779 & - & 8.466 &  &  \\ \hline\hline
\end{tabular}%
\end{table*}
\vspace{0.2cm}
\setlength{\abovecaptionskip}{0.3cm} \vspace{0.2cm}

\subsection{Triply heavy tetraquarks $sQ\bar{Q}\bar{Q\prime }$.}

The triply heavy tetraquarks $sQ\bar{Q}\bar{Q}$ consists of two types of
tetraquark states: the $sb\bar{c}\bar{b}$ and the $sc\bar{c}\bar{b}$. For
them, the $2^{+}$ state contains two states ($\phi _{2}\chi _{1}$, $\phi
_{1}\chi _{1}$) and the $1^{+}$ state forms six mixed states $c_{1}\phi
_{2}\chi _{2}+c_{2}\phi _{2}\chi _{4}+c_{3}\phi _{2}\chi _{5}+c_{4}\phi
_{1}\chi _{2}+c_{5}\phi _{1}\chi _{4}+c_{6}\phi _{1}\chi _{5}$ in the
subspace \{$\phi _{2}\chi _{2},\phi _{2}\chi _{4},\phi _{2}\chi _{5}$, $\phi
_{1}\chi _{2},\phi _{1}\chi _{4},\phi _{1}\chi _{5}$\}. Additionally, the $%
0^{+}$ state forms the mixed states, $c_{1}\phi _{2}\chi _{3}+c_{2}\phi
_{2}\chi _{6}+c_{3}\phi _{1}\chi _{3}+c_{4}\phi _{1}\chi _{6}$, in the
subspace \{$\phi _{2}\chi _{3},\phi _{2}\chi _{6},\phi _{1}\chi _{3},\phi
_{1}\chi _{6}$\}. \renewcommand{\tabcolsep}{0.6cm} \renewcommand{%
\arraystretch}{1.5}
\begin{table*}[tph]
\caption{Computed masses (GeV) triply heavy tetraquarks $sQ\bar{Q}\bar{Q}%
^{\prime }$, with bag radii  $R_{0}$ in GeV$^{-1}$. The masses are compared
to the other calculations. ${(sb\bar{c}\bar{b})}$ state's charge radii 
respectively are $0.54$(fm); ${(sb\bar{c}\bar{b})}$ state's Bag radii  $%
R_{0} $ respectively are $4.490$ is in GeV$^{-1}$. ${(sc\bar{c}\bar{b})}$
state's charge radii  respectively are $0.21$(fm). ${(sc\bar{c}\bar{b})}$
state's Bag radii  $R_{0}$ respectively are $4.75$ is in GeV$^{-1}$.}
\label{tab:3heavytetraquark4}%
\begin{tabular}{cccccccc}
\hline\hline
\textrm{State} & $J^{P}$ & \textrm{Eigenvector} & $M_{bag}$ & $\mu_{bag}$ &
\cite{Xin:2022} & \textrm{Scattering state} &  \\ \hline
${(sb\bar{c}\bar{b})}$ & ${2}^{+}$ & 1.00 & 12.149 & -1.12 & 11.558 &  &  \\
&  & 1.00 & 12.168 & -1.12 & 11.833 &  &  \\
& ${1}^{+}$ & (-0.29,0.31,-0.29,0.66,-0.37,-0.40) & 12.040 & -0.08 & 11.416
& $\bar{D}_{s}\Upsilon$ &  \\
&  & (-0.33,-0.35,-0.52,-0.34,-0.55,-0.29) & 12.107 & -1.12 & 11.496 &  &
\\
&  & (-0.57,0.21,0.41,-0.50,-0.10,0.45) & 12.134 & -0.95 & 11.549 &  &  \\
&  & (-0.06,-0.58,0.62,0.36,-0.37,-0.07) & 12.151 & -0.81 & 11.785 &  &  \\
&  & (-0.69,0.01,0.02,0.26,0.48,-0.49) & 12.166 & -0.35 & 11.803 &  &  \\
&  & (-0.10,-0.63,-0.30,0.03,0.43,0.56) & 12.168 & -0.06 & 11.834 &  &  \\
& ${0}^{+}$ & (-0.29,-0.43,0.82,0.25) & 12.029 & - & 11.355 & $\bar{D}%
_{s}^{*}\eta_{b}$ &  \\
&  & (-0.69,0.31,-0.26,0.60) & 12.097 & - & 11.539 & $B_{s}\bar{B}_{c}$ &
\\
&  & (0.21,0.84,0.51,0.02) & 12.144 & - & 11.761 &  &  \\
&  & (0.63,-0.13,-0.07,0.76) & 12.179 & - & 11.837 &  &  \\
${(sc\bar{c}\bar{b})}$ & ${2}^{+}$ & 1.00 & 8.831 & -0.57 & 8.432 & $\bar{D}%
_{s}^{*}$${B}_{c}^{*}$ &  \\
&  & 1.00 & 8.840 & -0.57 & 8.566 &  &  \\
& ${1}^{+}$ & (0.25,-0.22,0.60,-0.57,0.40,-0.22) & 8.720 & -0.14 & 8.288 & $%
\bar{D}_{s}{B}_{c}^{*}$ &  \\
&  & (-0.18,0.45,0.67,0.39,0.12,0.37) & 8.772 & -0.41 & 8.368 &  &  \\
&  & (-0.71,0.21,-0.07,-0.62,-0.11,0.22) & 8.799 & -0.63 & 8.408 &  &  \\
&  & (0.39,0.64,-0.38,-0.21,0.47,0.16 & 8.814 & -0.08 & 8.499 &  &  \\
&  & (-0.11,-0.31,-0.19,0.28,0.77,-0.05) & 8.837 & -0.13 & 8.531 &  &  \\
&  & (0.23,-0.46,-0.06,-0.10,0.03,0.86) & 8.860 & -0.33 & 8.551 &  &  \\
& ${0}^{+}$ & (-0.17,-0.63,0.75,0.13) & 8.692 & - & 8.220 & $\bar{D}_{s}{B}%
_{c}$ &  \\
&  & (0.81,-0.29,0.03,-0.52) & 8.762 & - & 8.375 &  &  \\
&  & (0.20,0.72,0.66,-0.06) & 8.799 & - & 8.488 &  &  \\
&  & (0.53,-0.04,-0.04,0.85) & 8.873 & - & 8.561 &  &  \\ \hline\hline
\end{tabular}%
\end{table*}
\vspace{0.2cm}
\begin{figure}[th]
\setlength{\abovecaptionskip}{0.3cm}
\par
\begin{center}
\includegraphics[width=0.5\textwidth]{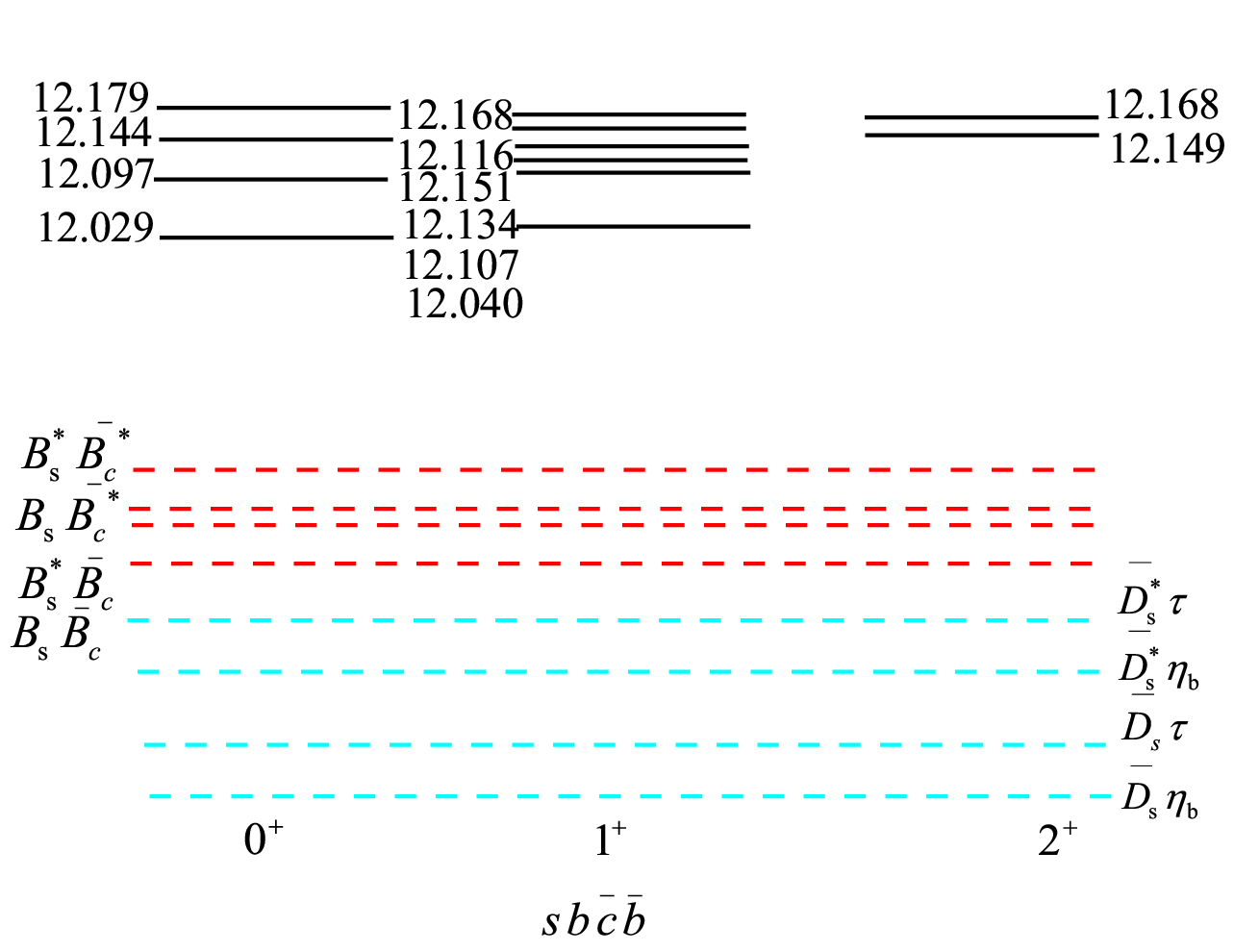} \includegraphics[width=0.5%
\textwidth]{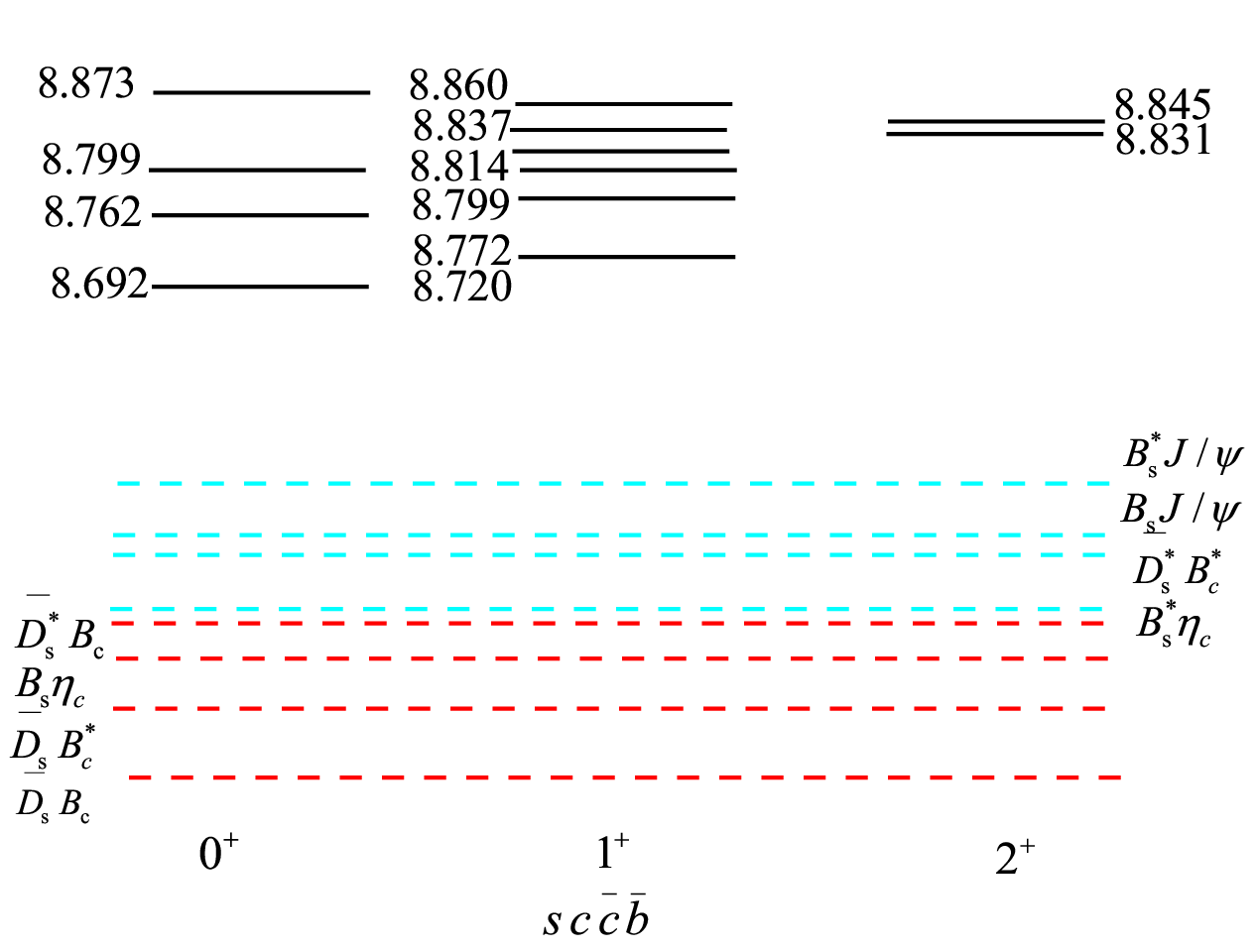}
\end{center}
\caption{Mass spectra(GeV, the sold lines) of the tetraquarks $sb\bar{c}\bar{%
b}$ and $sc\bar{c}\bar{b}$, with the meson-meson thresholds (GeV) plotted in
dotted lines. }
\label{fig:6}
\end{figure}
The computed masses and other properties of the tetraquark states are listed
in Table \ref{tab:3heavytetraquark4}. With their masses exceeding the
two-meson thresholds, the triply heavy tetraquark $sc\bar{c}\bar{b}$ can
decay to the channels $B_{s}^{\ast }J/\psi $, $B_{s}J/\psi $, $\bar{D}%
_{s}^{\ast }B_{c}^{\ast }$, $B_{s}^{\ast }\eta _{c}$, $\bar{D}_{s}^{\ast
}B_{c}$, $B_{s}\eta {c}$, $\bar{D}_{s}B_{c}^{\ast }$ and $\bar{D}_{s}\bar{B}%
_{c}$, as depicted in FIG. \ref{fig:6}. The state $sb\bar{c}\bar{b}$ can
decay to $B_{s}^{\ast }\bar{B}_{c}^{\ast }$, $B_{s}^{\ast }\bar{B}_{c}$, $%
B_{s}\bar{B}_{c}$, $B_{s}\bar{B}_{c}^{\ast }$, $\bar{D}_{s}^{\ast }\Upsilon $%
, $\bar{D}_{s}^{\ast }\eta {b}$, $\bar{D}_{s}\Upsilon $ or $\bar{D}_{s}\eta
_{b}$. Both of the tetraquarks $sc\bar{c}\bar{b}$ and $sb\bar{c}\bar{b}$ can
be the candidates of the scattering states, as noted in Table. \ref%
{tab:3heavytetraquark4}.

\section{Summary}

\label{Discussions} In this work, we employ the MIT bag model which
incorporates chromomagnetic interactions and enhanced binding energy to
systematically study the masses and other properties of all triply heavy
tetraquarks ($qQ_{1}\bar{Q}_{2}\bar{Q_{3}}$) in their ground states. The
variational analysis is applied to compute the masses, magnetic moments and
charge radii of all strange and nonstrange (ground) states of triply heavy
tetraquarks. Our computation indicates that all of these triply heavy
tetraquark states are above their two-meson thresholds and can decay to
their two-meson final states via strong interaction.
We observe that the tetraquark states in configuration $6_{c}\otimes \bar{6}%
_{c}$ are usually heavier than that in $\bar{3}_{c}\otimes {3}_{c}$ about $%
50-150$ MeV for the same quantum number $J^{P}$,  with exception for all $J^{P}=1^{+}$ states with one
multiplet in the middle of two others in the sense of the mass order. Furthermore, the mass splitting between them
becomes smaller when more bottom quarks are involved.
By the way, we estimate the relative decay widths of two-meson final states
of the triply heavy tetraquarks and predict the primary decay channels for
each of them. We hope that our prediction helps to find the triply heavy
tetraquarks during experimental search in the future.

There are some scattering states among the tetraquark states ($sQ\bar{Q}\bar{%
Q\prime }$, $nQ\bar{Q}\bar{Q\prime }$) which is based on the numerical
evaluation of the $I$-th decay channel weight ${\lvert c_{I}\rvert }^{2}$.
When ${\lvert c_{I}\rvert }^{2}$ tends to be close to unity, the tetraquark
state turns out to be a scattering state, for which the decay and width
computation are considered.

\textbf{Acknowledgments }

D. J thanks Xue-Qian Li for discussions. D. J is supported by the National
Natural Science Foundation of China under the no. 12165017.

\appendix
\section{}

In Eq. (\ref{Iij}) of the MIT bag model, the rational function $%
F(x_{i},x_{j})$ is given by \cite{DeGrand:1975}.
\begin{equation}
\begin{aligned} F(x_{i}, x_{j})={\left(x_{i} {\rm
sin}^{2}x_{i}-\frac{3}{2}y_{i}\right)}^{-1} {\left(x_{j} {\rm
sin^{2}}x_{j}-\frac{3}{2}y_{j}\right)}^{-1} \\
\left\{-\frac{3}{2}y_{i}y_{j}-2x_{i}x_{j} {\rm sin}^{2}x_{i} {\rm
sin}^{2}x_{j} +\frac{1}{2}x_{i}x_{j}\left[2x_{i} {\rm Si}(2x_{i})
\right.\right.\\ \left.\left. +2x_{j} {\rm Si}(2x_{j}) -(x_{i}+x_{j}) {\rm
Si}(2(x_{i}+x_{j})) \right.\right.\\ \left.\left. -(x_{i}-x_{j}) {\rm
Si}(2(x_{i}-x_{j})) \right] \right\}, \end{aligned}  \label{F}
\end{equation}%
wtih $y_{i}=x_{i}-\mathrm{cos}(x_{i})\mathrm{sin}(x_{i})$. Here, the running
strong coupling $\alpha _{s}(R)$ is defined as \cite{He:2004},
\begin{equation}
\alpha _{s}(R)=\frac{0.296}{\mathrm{ln}\left[ 1+{\left( R(0.281\text{
GeV\thinspace })\right) }^{-1}\right] },  \label{alphaS-mine}
\end{equation}%
in which the bag radius $R$ depends upon a dimensionless parameter $%
x_{i}=x_{i}(mR)$ via a transcendental equation.
\begin{equation}
\tan x_{i}=\frac{x_{i}}{1-m_{i}R-\left( m_{i}^{2}R^{2}+x_{i}^{2}\right)
^{1/2}}.  \label{transc}
\end{equation}

\section{}

The matrix elements in the CMI (\ref{CMI}) can be evaluated via the
following relations
\begin{equation}
{\left\langle \boldsymbol{\lambda }_{i}\cdot \boldsymbol{\lambda }%
_{j}\right\rangle }_{nm}=\sum_{\alpha =1}^{8}Tr\left( c_{in}^{\dagger
}\lambda ^{\alpha }c_{im}\right) Tr\left( c_{jn}^{\dagger }\lambda ^{\alpha
}c_{jm}\right) ,  \label{colorfc}
\end{equation}%
\begin{equation}
{\left\langle \boldsymbol{\sigma }_{i}\cdot \boldsymbol{\sigma }%
_{j}\right\rangle }_{xy}=\sum_{\alpha =1}^{3}Tr\left( \chi _{ix}^{\dagger
}\sigma ^{\alpha }\chi _{iy}\right) Tr\left( \chi _{jx}^{\dagger }\sigma
^{\alpha }\chi _{jy}\right) ,  \label{spinfc}
\end{equation}%
where $(n,m)$ and $(x,y)$ refer to the indices of the color and spin states,
respectively, $i$ and $j$ refer to quarks or antiquarks. Here, $c_{in}$
stand for the color bases (color wavefunctions) of the quark(antiquark) $i$,
and $\chi _{ix}$ stand for the spin bases of the quark(antiquark) $i$.

Based on formulas (\ref{colorfc}) and the color wavefunctions in subsection
2.B, one can compute all color factors for the normal hadrons (mesons and
baryons) addressed in this work. For instance, the color factors is
\begin{equation}
\begin{aligned} &\left\langle \boldsymbol{\lambda_{1}}\cdot
\boldsymbol{\lambda_{2}} \right\rangle=-\frac{16}{3}, \end{aligned}
\label{cfcT}
\end{equation}%
for the quark-antiquark in meson with the wave function $(\phi ^{M})$, and
the color factor is

\begin{equation}
\begin{aligned} &\left\langle \boldsymbol{\lambda_{1}}\cdot
\boldsymbol{\lambda_{2}} \right\rangle=\left\langle
\boldsymbol{\lambda_{1}}\cdot \boldsymbol{\lambda_{3}}
\right\rangle=\left\langle \boldsymbol{\lambda_{2}}\cdot
\boldsymbol{\lambda_{3}} \right\rangle=-\frac{8}{3}, \end{aligned}
\label{cfcT}
\end{equation}%
for quark pairs in the baryon with $(\phi ^{B})$.

For tetraquarks with two color configurations $(\phi _{1}^{B},\phi _{2}^{B})$
in the zero-order approximation, the color factors for them are

\begin{equation}
\begin{aligned} &\left\langle \boldsymbol{\lambda_{1}}\cdot
\boldsymbol{\lambda_{2}} \right\rangle= \left\langle
\boldsymbol{\lambda_{3}}\cdot \boldsymbol{\lambda_{4}} \right\rangle=
\begin{bmatrix} \frac{4}{3} & 0 \\ 0 & -\frac{8}{3} \end{bmatrix}, \\
&\left\langle \boldsymbol{\lambda_{1}}\cdot \boldsymbol{\lambda_{3}}
\right\rangle= \left\langle \boldsymbol{\lambda_{2}}\cdot
\boldsymbol{\lambda_{4}} \right\rangle= \begin{bmatrix} -\frac{10}{3} &
2\sqrt{2} \\ 2\sqrt{2} & -\frac{4}{3} \end{bmatrix}, \\ &\left\langle
\boldsymbol{\lambda_{1}}\cdot \boldsymbol{\lambda_{4}} \right\rangle=
\left\langle \boldsymbol{\lambda_{2}}\cdot \boldsymbol{\lambda_{3}}
\right\rangle= \begin{bmatrix} -\frac{10}{3} & -2\sqrt{2} \\ -2\sqrt{2} &
-\frac{4}{3} \end{bmatrix}, \end{aligned}  \label{cfcT}
\end{equation}%
which are $2\times 2$ matrices in the subspace spanded by the base
wavefunctions ($\phi _{1}^{T},\phi _{2}^{T}$). Here, base $\phi _{1}^{T}$
corresponds to the $\boldsymbol{6}_{c}\otimes \boldsymbol{\bar{6}}_{c}$
color configuration while $\phi _{2}^{T}$ corresponds to the $\boldsymbol{%
\bar{3}}_{c}\otimes \boldsymbol{3}_{c}$ color configuration.


\begin{thebibliography}{99}
\bibitem{PDG:2020} R.~L.~Workman et al. [Particle Data Group], Review of
Particle Physics, PTEP 2022, 083C01 (2022).

\bibitem{LhcXiLf:prl18} M. Ablikim et al., Confirmation of a charged
charmoniumlike state, Phys. Rev. D, 092006(2015), arXiv:1509.01398[hep-ex].

\bibitem{M. Ablikim et al:2014} M. Ablikim et al, Observation of a Charged
Charmoniumlike Structure, Phys. Rev. Lett. \textbf{112}, 132001 (2014),
arXiv:1308.2760 [hep-ex]

\bibitem{M. Ablikim:2013} M. Ablikim et al, Observation of a charged
charmoniumlike structure, 242001 (2013), arXiv:1309.1896 [hep-ex]

\bibitem{M. Ablikim et al:2021} M. Ablikim et al, Observation of a
near-threshold structure in the ${K}^{+}$ recoil-mass Spectra, Phys. Rev.
Lett. \textbf{126},102001(2021)

\bibitem{R. Aaij et al:2021} R. Aaij et al, Observation of new resonances
decaying, Phys. Rev. Lett. \textbf{127}, 082001 (2021), arXiv:2103.01803
[hep-ex]

1\bibitem{Belle:2014nuw}
K.~Chilikin \textit{et al.} [Belle],
Phys. Rev. D \textbf{90} (2014) no.11, 112009
[arXiv:1408.6457 [hep-ex]].

2\bibitem{Belle:2013shl} K.~Chilikin \textit{et al.} [Belle], Experimental
constraints on the spin and parity of the \$Z\$(4430)\$\symbol{94}+\$, Phys.
Rev. D \textbf{88} (2013) 074026 [arXiv:1306.4894 [hep-ex]].

3\bibitem{Yan:2021tcp} M.~J.~Yan, F.~Z.~Peng, M.~S\'{a}nchez S\'{a}nchez and
M.~Pavon Valderrama, Axial meson exchange and the \$Z\_c(3900)\$ and
\$Z\_\{cs\}(3985)\$ resonances as heavy hadron molecules, Phys. Rev. D
\textbf{104} (2021)114025[arXiv:2102.13058 [hep-ph]].

4\bibitem{Deng:2021gnb}
C.~Deng and S.~L.~Zhu,
Phys. Rev. D \textbf{105}, no.5, 054015 (2022)
[arXiv:2112.12472 [hep-ph]].

5\bibitem{BESIII:2013ris}
M.~Ablikim \textit{et al.} [BESIII],
Phys. Rev. Lett. \textbf{110}, 252001 (2013)
[arXiv:1303.5949 [hep-ex]].

6\bibitem{Belle:2013yex}
Z.~Q.~Liu \textit{et al.} [Belle],
Phys. Rev. Lett. \textbf{110}, 252002 (2013)
[erratum: Phys. Rev. Lett. \textbf{111}, 019901 (2013)]
[arXiv:1304.0121 [hep-ex]].

7\bibitem{LHCb:2020bwg} R.~Aaij \textit{et al.} [LHCb],
Sci. Bull. \textbf{65} (2020) no.23, 1983-1993
[arXiv:2006.16957 [hep-ex]].

14\bibitem{TccPoly:2021} I. Polyakov, [on behalf of LHCb Collaboration], Talk
at the Euro. Phys. Soc. Conference on High Energy Physics, 29 July (2021).

15\bibitem{L.Maiani:2005} L. Maiani and F. Piccinini and A. D. Polosa and V.
Riquer,Diquark-antidiquark states with hidden or open charm , Physical
Review D. \textbf{71}, 014028 (2005), arXiv:hep-ph/0412098 [hep-ph]

16\bibitem{Woosung Park:2014} Woosung Park and Su Houng Lee,Color spin wave
functions of heavy tetraquark states , Nuclear Physics A. \textbf{925},
02.008 (2014), arXiv:1311.5330 [hep-ph]

17\bibitem{Muhammad Naeem:2018} Muhammad Naeem Anwar and Jacopo Ferretti and
Elena Santopinto,Spectroscopy of the hidden-charm , Physical Review D
\textbf{98}, 094015 (2018), arXiv:1805.06276 [hep-ph]

18\bibitem{Nils A.:1994} Nils A.,From the deuteron to deusons, an analysis of
deuteronlike meson-meson bound states, Zeitschrift Physik C Particles and
Fields \textbf{61}, (1994), arXiv:1707.09575 [hep-ph]

19\bibitem{Nils A Trnqvist:2004} Nils A Trnqvist,Heavy-quark symmetry implies
stable heavy tetraquark mesons $Q_{i}Q_{j}\bar{q}_{k}\bar{q}_{l}$, Physics
Letters B \textbf{590}, (2004), arXiv:hep-ph/0402237 [hep-ph]

20\bibitem{Eric S. Swanson:2004} Eric S. Swanson,Short range structure in the
X(3872), Physics Letters B \textbf{588}, (2004), arXiv:hep-ph/0311229
[hep-ph]

21\bibitem{C. Hanhart:2007} C. Hanhart and Yu. S. Kalashnikova and A. E.
Kudryavtsev and A. V. Nefediev, Physical Review D. \textbf{76}, 034007
(2007), arXiv:0704.0605 [hep-ph]

22\bibitem{F. Aceti:2012} F. Aceti and R. Molina and E. Oset, Physical Review
D. \textbf{86}, 113007 (2012), arXiv:1207.2832 [hep-ph]

23\bibitem{Rui Chen:2016} Rui Chen and Xiang Liu and Yan-Rui Liu and Shi-Lin
Zhu,Predictions of the hidden-charm molecular states with the four quark
components, The European Physical Journal C. \textbf{76}, (2016),
arXiv:1511.03439 [hep-ph]

24\bibitem{Shi-Lin Zhu:2005} Shi-Lin Zhu,The possible interpretations of
Y(4260), Physics Letters B. \textbf{625}, (2005), arXiv:hep-ph/0507025
[hep-ph]

25\bibitem{A. Esposito:2016} A. Esposito and A. Pilloni and A.D.
Polosa,Hybridized tetraquarks, Physics Letters B. \textbf{758}, (2016),
arXiv:1603.07667 [hep-ph]

26\bibitem{Karliner:2017} Marek Karliner and Jonathan L. Rosner,Quark-level
analogue of nuclear fusion with doubly heavy baryons, Phys. Rev. Lett.
\textbf{551}, 24289(2017),arXiv:1708.02547 [hep-ph]

27\bibitem{Eichten:2017} E.~J.~Eichten and C.~Quigg,Heavy-quark symmetry
implies stable heavy tetraquark mesons $Q_{i}Q_{j}\bar{q}_{k}\bar{q}_{l}$,
Phys. Rev. Lett. \textbf{119}, 202002 (2017), arXiv:1707.09575 [hep-ph]

28\bibitem{Meng-Lin Du :2013} Meng-Lin Du and Wei Chen and Xiao-Lin Chen and
Shi-Lin Zhu, Physical Review D. \textbf{87}, 014003 (2013), arXiv:1209.5134
[hep-ph]

29\bibitem{Fleck:1989} S.~Fleck and J.~M.~Richard, Baryons with double charm,
Prog. Theor. Phys. \textbf{82}, 760-774 (1989)

30\bibitem{Ebert:2004} D.~Ebert, R.~N.~Faustov, V.~O.~Galkin and
A.~P.~Martynenko,Semileptonic decays of doubly heavy baryons in the
relativistic quark model,Phys. Rev. D \textbf{70}, 014018 (2004) [erratum:
Phys. Rev. D \textbf{77}, 079903 (2008)], arXiv:hep-ph/0404280 [hep-ph]

31\bibitem{Roberts:2007} W.~Roberts and M.~Pervin, Heavy baryons in a quark
model, Int. J. Mod. Phys. A \textbf{23}, 2817-2860 (2008), arXiv:0711.2492
[nucl-th]

32\bibitem{Albertus:2006} C.~Albertus, E.~Hernandez, J.~Nieves and
J.~M.~Verde-Velasco, Static properties and semileptonic decays of doubly
heavy baryons in a nonrelativistic quark model, Eur. Phys. J. A \textbf{32},
183-199 (2007) [erratum: Eur. Phys. J. A \textbf{36}, 119 (2008)],
arXiv:hep-ph/0610030 [hep-ph]

33\bibitem{Giannuzzi:2009} F.~Giannuzzi, Doubly heavy baryons in a Salpeter
model with AdS/QCD inspired potential, Phys. Rev. D \textbf{79}, 094002
(2009), arXiv:0902.4624 [hep-ph]

34\bibitem{Bernotas:2008} A.~Bernotas and V.~Simonis, Mixing of heav baryons
in the bag model calculations, Lith. J. Phys. Tech. Sci. \textbf{48}, 127
(2008), arXiv:0801.3570 [hep-ph]

35\bibitem{Liu:2018} M.~Z.~Liu, Y.~Xiao and L.~S.~Geng, Magnetic moments of
the spin-1/2 doubly charmed baryons in covariant baryon chiral perturbation
theory, Phys. Rev. D \textbf{98}, 014040 (2018), arXiv:1807.00912 [hep-ph]

36\bibitem{He:2004} D.~H.~He, K.~Qian, Y.~B.~Ding, X.~Q.~Li and P.~N.~Shen,
Evaluation of spectra of baryons containing two heavy quarks in bag model,
Phys. Rev. D \textbf{70}, 094004 (2004), arXiv:0403301[hep-ph]

37\bibitem{KR:2014} M.~Karliner and J.~L.~Rosner, Baryons with two heavy
quarks: Masses, production, decays, and detection, Phys. Rev. D \textbf{90},
094007 (2014), arXiv:1408.5877 [hep-ph]

38\bibitem{Mutuk:2023yev}
H.~Mutuk,
Eur. Phys. J. C \textbf{83} (2023) no.5, 358
doi:10.1140/epjc/s10052-023-11526-7
[arXiv:2305.03358 [hep-ph]].

39\bibitem{Meng:2023jqk}
L.~Meng, Y.~K.~Chen, Y.~Ma and S.~L.~Zhu,
Phys. Rev. D \textbf{108} (2023) no.11, 114016
doi:10.1103/PhysRevD.108.114016
[arXiv:2310.13354 [hep-ph]].

40\bibitem{Liu:2022jdl}
X.~Liu, Y.~Tan, D.~Chen, H.~Huang and J.~Ping,
Phys. Rev. D \textbf{107} (2023) no.5, 054019
doi:10.1103/PhysRevD.107.054019
[arXiv:2205.08281 [hep-ph]].

41\bibitem{Lu:2021kut}
Q.~F.~L\"u, D.~Y.~Chen, Y.~B.~Dong and E.~Santopinto,
Phys. Rev. D \textbf{104} (2021) no.5, 054026
doi:10.1103/PhysRevD.104.054026
[arXiv:2107.13930 [hep-ph]].

42\bibitem{Guo:2021yws}
T.~Guo, J.~Li, J.~Zhao and L.~He,
Phys. Rev. D \textbf{105} (2022) no.1, 014021
doi:10.1103/PhysRevD.105.014021
[arXiv:2108.10462 [hep-ph]].

43\bibitem{Cui:2006mp}
Y.~Cui, X.~L.~Chen, W.~Z.~Deng and S.~L.~Zhu,
HEPNP \textbf{31} (2007), 7-13
[arXiv:hep-ph/0607226 [hep-ph]].







44\bibitem{DeGrand:1975} T.~A.~DeGrand, R.~L.~Jaffe, K.~Johnson and
J.~E.~Kiskis, Masses and other parameters of the light hadrons, Phys. Rev. D
\textbf{12}, 2060 (1975)

45\bibitem{Zhang:2021} Wen-Xuan Zhang and Hao Xu and Duojie Jia, Masses and
magnetic moments of hadrons with one and two open heavy quarks: Heavy
baryons and tetraquarks, Physical Review D. \textbf{104},
114011(2021),arXiv:2109.07040[hep-ph].

46\bibitem{Xin:2022} Xin-Zhen Weng and Wei-Zhen Deng and Shi-Lin Zhu, Triply
heavy tetraquark states, Physical Review D. \textbf{105}, 034026
(2022),arXiv:2109.05243[hep-ph].

47\bibitem{Chodos:1974} A.~Chodos, R.~L.~Jaffe, K.~Johnson and C.~B.~Thorn,
Baryon structure in the bag theory, Phys. Rev. D \textbf{10}, 2599 (1974)
\end{thebibliography}
\end{document}